\documentclass[prb,twocolumn,amsmath,amssymb,aps,superscriptaddress,eqsecnum]{revtex4-1}
\usepackage{graphicx}
\usepackage{dcolumn}
\usepackage{bm}
\begin{document}

\title{Decorated defect condensate, a window to unconventional quantum phase transitions in Weyl semimetals}

\author{Yizhi You}
\affiliation{Department of Physics and Institute for Condensed Matter Theory, University of Illinois at Urbana-Champaign, 
Illinois 61801}

\date{\today}
\begin{abstract}
We investigate the unconventional quantum phase transitions in Weyl semimetals. The emergent boson fields, coupling with the Weyl fermion bilinears, contain a Wess-Zumino-Witten term or topological $\Theta$ term inherited from the momentum space monopoles carried by Weyl points. 
Three types of unconventional quantum critical points will be studied in order: (1) The transition between two distinct symmetry breaking phases whose criticality is beyond Landau's paradigm.
 (2) The transition between a symmetry breaking state to a topological ordered state.  (3) The transition between $3d$ topological order phase to trivial disordered phase whose criticality could be traced back to a $Z_2$ symmetry breaking transition in $4d$.
  The essence of these unconventional critical points lies in the fact that the topological defect of an order parameter carries either a nontrivial quantum number or a topological term so the condensation of the defects would either break some symmetry or give rise to a topological order phase with nontrivial braiding statistics.

\end{abstract}

\maketitle

\section{Introduction and Motivation}
Throughout the past decades, the Landau-Ginzburg-Wilson(LGW) theory successfully describes a large class of continuous phase transition in terms of fluctuating order parameters. In addition to the phase transition appearing at finite temperature driven by thermal fluctuation, the quantum phase transitions, controlled by quantum fluctuations in terms of external parameters at zero temperature, can also be explained via the theoretical framework of LGW paradigm. 

While the condensation of an order parameter drives the system into an ordered phase, the order to disorder transition, on the opposite trend could be regarded as the condensation of order parameter defects. To enumerate, the transition between Ising ferromagnet phase to  paramagnetic phase can be realized by condensation of domain walls\cite{pfeuty1970one}; the superfluid to Mott insulator transition can be driven by superfluid vortex condensate\cite{fisher1989boson,lee1990boson}.   

However, there still appears some unconventional quantum critical points which are beyond the Landau-Ginzburg-Wilson(LGW) type\cite{senthil2005deconfined,senthil2004deconfined,vishwanath2004quantum,senthil2006competing,grover2008topological,moon2012skyrmions,tsui2015quantum}. One of the explicit examples is the deconfined quantum criticality between VBS/nematic order to Neel order in frustrated spin systems\cite{senthil2005deconfined,senthil2004deconfined,vishwanath2004quantum,senthil2006competing,levin2004deconfined,nahum2015emergent}, where the competing orders between different symmetry broken states are connected by a continuous transition with only one relevant coupling constant. Such quantum criticality cannot be described by LGW formalism which suggests two distinct ordered phase shall either be connected by an intermediate phase or experience a first order transition. 

In addition, the discovery of topological matter\cite{wen1990topological,laughlin1983anomalous} introduces new species of matter which cannot be probed by any local operators. Consequently, the transitions among topological matter, symmetry protected topological order and trivial phase are also beyond LGW formalism as they share the same symmetry and become identical within any local probe\cite{burnell2012phase,fisher1990quantum,xu2012unconventional,lee1991anyon}. In the absence of local order parameter, one cannot plainly, follow the LGW approach to described the phase transition in terms of order parameter fluctuations. The critical theory between topological order(or SPT) to trivial phase is studied by a bundle of pioneers\cite{senthil2014symmetry,chen2014symmetry,you2014symmetry,fidkowski2011topological,moon2012exotic}. The common wisdom of these approach involves adding a topological term in addition to the conventional LGW theory. The topological $\Theta$ term in addition to the classical NL$\sigma$M provides a new critical point which exactly describes the critical theory between SPT to trivial phase\cite{senthil2014symmetry,you2014symmetry}.  In addition, the transition from topological order to trivial phase can be approached by anyon condensate where the anyons, as a Lagrangian subgroup of the topological excitations, would confine any other topological quasiparticle after its proliferation\cite{burnell2012phase,kong2014anyon,barkeshli2010anyon,metlitski2014interaction}.

Apart from the unconventional quantum critical points beyond the LGW formalism, there also exist some exotic quantum criticality as an enriched LGW theory, $e.g.$, the phase transition between the symmetry breaking state to a topological state(or symmetry protected topological order state)\cite{xu2012unconventional,you2016stripe,tsui2015quantum,you2016geometry,cho2015condensation}.  Such transition, connecting order to disorder phases, is still within the LGW paradigm. However, the defect of the order parameter in the symmetry breaking phase is decorated with some topological terms\cite{xu2012unconventional,chen2013critical,tsui2015topological,you2016stripe,chen2014symmetry}. Accordingly, if one proliferates the defect to disordered the phase, the coupling between defect and topological term in the effective theory makes itself distinguish from a trivial disordered phase, namely we obtain a topological state(or symmetry protected topological order state).

In this paper, we investigate three types of unconventional transitions in Weyl semimetal systems at three spatial dimensions. 1) The transition between two distinct symmetry  breaking phases. 2) The transition between  symmetry  breaking state and  topological ordered state in $3+1 d$. 3) The transition between  topological order state and  trivial(or SPT) state. The essence of these unconventional criticality is to decorate the topological defect with some nontrivial quantum number or topological terms\cite{you2016stripe,chen2014symmetry}. Our starting point is the Weyl semimetals\cite{wan2011topological,zyuzin2012topological} who contain eight Weyl cones coupling with some fluctuating boson variables. Owing to the momentum space monopole carried by each Weyl point, the effective theory of the boson variables forms a Wess-Zumino-Witten(WZW) term and the topological defect of the boson field therefore carries some nontrivial quantum number or topological term. 

The outlines of our paper are organized as follow. 
In section II, we review the deconfined quantum criticality in $2d$ Mott insulators. In Section III, we investigate the SU(2) soliton condensation towards a charge superfluid state. We couple the SU(2) field to the Weyl semimetal where the SU(2) symmetry breaking phase corresponds to the chiral symmetry breaking of  $QED_4$. The soliton of the SU(2) degree of freedom carries charge $2e$ and its proliferation restores the SU(2) symmetry but meanwhile breaks the charge U(1) symmetry towards a superconductivity order. Such charged soliton condensation transition connecting different symmetry broken state is also beyond the LGW type. The essence of this exotic quantum phase transition arises from the nontrivial quantum number carried by the order parameter defect and the criticality therefore contains 
an emergent WZW term.   

In Section IV, we study the transition between symmetry broken state and $3d$ topological ordered state.  We start with the $3d$ pair density wave(PDW) state where the fermions in each Weyl cone form an s-wave pairing and condense with a global momentum. The nodal plane of the PDW state contains gapless modes. We couple the gapless fermions with an O(3) rotor and turn the gapless nodal plane into a gapped state whose effective theory is equivalent to a $2d$ topological paramagnetic phase\cite{you2016stripe}. After we condense the dislocation and disclination to disorder the PDW order, the ground state(GS) wave function can be written in terms of the superposition of all close nodal membranes decorated with a topological paramagnetic state. Alternatively, if we express the GS wave function in terms of the O(3) rotor degree of freedom, the wave function is the condensation of skyrmion flux loops decorated with fluctuating domain walls.
The open membrane, whose boundary contains a half superconducting(SC) vortex loop is a deconfined loop excitation. Meanwhile, the end point of the flux line contains a monopole of the O(3) rotor as a deconfined particle excitation. The monopole has $\pi$ statistics with the  half SC vortex and the system is therefore in a $3d$ $Z_2$ topological order phase equivalent to the $3d$ toric code model\cite{yu2008topological,moradi2015universal}. The spirit of this unconventional transition lies in the fact that the effective theory of the nodal plane, as a decorated topological paramagnetic state, contains a topological $\Theta$ term.

In Section V, we focus on the transition between SPT state and topological ordered state in $3d$ from interacting Weyl semimetals. We look into the SPT phases whose surface contain topological order and let the surface topological order saturates into the bulk by domain wall proliferation. The proliferation of domain walls decorated with an anomalous $2d$ topological order drives the system into a $3d$ topological order phase. On the opposite trend, the transition from the topological order to an SPT phase can be realized by loop condensate which confines other topological excitations. 
Finally, we also map the transition between SPT and topological order phase in $3d$ to a spontaneously $Z_2$ symmetry breaking transition in $4d$. This scenery generates a connection between topological and LGW type transition in different dimensions from a holographic view.

\section{Novel Quantum Criticality between two symmetry broken phases}

A large class of spontaneous symmetry breaking phase transitions can be described by the Landau-Ginzburg-Wilson(LGW) theory. The LGW paradigm demonstrates that a continuous phase transition occurs from a symmetry broken phase to a disordered phase or vice versa. In addition, for two phase of matters with distinct symmetry breaking, the transition between them shall encounter with an intermediate phase which, both or neither symmetry are broken.

To illustrate, imagine we have a classical $O(M+N)$ rotor $\vec{n}$ described by the non-linear sigma model(NL$\sigma$M). 
\begin{align} 
&\mathcal{L}= \frac{1}{g} (\partial_{\mu} n_i)^2
\end{align}
The coupling constant $g$ tunes the fluctuation strength of the rotor. When $g$ is small, the rotor is in the ordered phase which breaks the rotation symmetry and contains $M+N-1$ gapless Goldstone mode. When $g$ is large, the theory is in the gapped disordered phase.
Now assume we slightly break the rotor from $O(M+N)$ to $O(M)\times O(N)$.
\begin{align} 
&\mathcal{L}= \frac{1}{g} \sum^{M}_{i=1} (\partial_{\mu} n_i)^2+\sum^{M+N}_{j=M+1}\frac{1}{g'} \sum(\partial_{\mu} n_j)^2
\end{align}
We then have two coupling constant $g$ and $g'$ which tunes the fluctuation strength of two rotors. The phase diagram therefore contains four distinct phases, a) $O(N)$ symmetry  breaking, b) $O(M)$ symmetry breaking, c) both symmetries are broken, d) both rotors are disordered. The phase transition between the $O(N)$ to $O(M)$ symmetry broken phase has to go through an intermediate region, where both symmetries were broken or both rotors were disordered. Else, the two different symmetry broken phases can also be connected by a multicritical point.

It was long recognized that two distinct symmetry broken phase cannot be connected by a continuous transition with only one relevant coupling constant. However, Senthil $et$ $al.$ \cite{senthil2004deconfined,senthil2005deconfined} proposed an exotic deconfined quantum critical theory beyond LGW paradigm. Such transition\cite{senthil2004deconfined,senthil2005deconfined} connects different symmetry breaking matter by a continuous transition, and the critical region is controlled by the emergent WZW term\cite{senthil2004deconfined,senthil2005deconfined}. The WZW term in the criticality suggests the defect of an order parameter carry some nontrivial quantum number. As a result, the condensation of symmetry defects restores the symmetry, but meanwhile breaks another symmetry associate with the quantum number. Before we proceed, let us first review the deconfined quantum phase transition in $2d$.

\subsection{Phase transition between Neel and VBS order in Mott insulator} 
The first prominent example on deconfined quantum criticality is discovered by Senthil $et$ $al.$ \cite{senthil2014symmetry,chen2014symmetry,you2014symmetry,fidkowski2011topological} in Mott insulator on  square lattice.  Deep in the Mott phase, the charge degree of freedom is frozen while the spin degree of freedom suffers from a variety of competing orders. 

In the Neel ordered phase, the spin $1/2$ on each cite forms an AF order which breaks the spin rotation symmetry. The path integral description of the quantum spin fluctuation is determined by the nonlinear sigma model with a BerryÕs phase factor\cite{senthil2006competing}. The Berry phase indicates the fact that the instanton event\cite{haldane1983continuum}  contributes a phase factor of $e^{i\pi/2}$ which enters into the path integral over all possible spacetime spin configurations. Accordingly, a single (spacetime)hedgehog cannot appear in the critical theory. Meanwhile, the instanton event for adding skyrmion quadrupole(with four skyrmion) does not generate any Berry phase with sign frustration in the path integral. However, the skyrmion quadrupole carries lattice momentum and the condensation of skyrmion restores the rotation symmetry but meanwhile breaks the translation symmetry by lattice momentum condensation. The four-fold hedgehog operator is dangerously irrelevant so the critical point is stable against that. However, as long as the skyrmion condense, the four-fold hedgehog becomes relevant and the
proliferation of instantons confines the U(1) gauge field carried by the original spinon. The resultant phase after skyrmion quadrupole condensate is a VBS state with dimerized spin order\cite{senthil2005deconfined,senthil2004deconfined,vishwanath2004quantum,senthil2006competing}. The VBS state restores the spin rotation symmetry but breaks the translation symmetry.

Alternatively, one can also approach the criticality from the VBS side. The VBS state on the square lattice breaks translation symmetry and there are 4 distinct VBS configurations describing a discrete $Z_4$ clock order parameter. The $Z_4$ vortex of such order parameter carries a spinon. The condensation of  $Z_4$ vortex, together with the spinon restores the translation symmetry but meanwhile breaks the O(3) rotation symmetry\cite{levin2004deconfined}. The quantum critical region is characterized by the O(5) WZW model where the O(3) spinon together with the O(2) clock order parameter forms an emergent O(5) rotor. (When approaching the criticality, the instantons corresponding to the 4-fold anisotropy are (dangerously) irrelevant so one can enlarge the symmetry from $Z_4$ to $U(1)$.)

 \section{Quantum phase transition in $3d$ connecting distinct symmetry breaking states}
The concrete example of the continuous transition between two phases with different broken symmetries was found in miscellaneous systems in $2d$ with strong interaction\cite{moon2012skyrmions,grover2008topological}, including frustrated magnets, bilayer graphene, etc. However, such mechanism and concrete examples in $3d$ is less explored as correlation effect is suppressed in higher dimension. Moon\cite{moon2015competing} demonstrated that with the presence of anomalies and relevant symmetry breaking operators, the Wess-Zumino-Witten model in higher dimension can precisely characterize similar unconventional quantum criticality.
 
In this part, we intend to construct a microscopic model as a platform for novel quantum phase transition between different symmetry breaking phases. As is pointed out by several pioneers\cite{moon2015competing,xu2013nonperturbative,xu2012unconventional,chan2016effective}, a WZW theory at the stable fixed point is essential in the critical theory connecting different symmetry breaking states since the WZW term itself connects an order parameter defect with another quantum number.  As a result, in order to acquire such phase transition beyond LGW paradigm, it is essential for us to start from a critical theory containing a WZW term. 

For deconfined quantum criticality in $2d$, the WZW term\cite{nahum2015emergent} emerges due to the special structure of the VBS vortex who contains spin $1/2$ degree of freedom in the vortex core. However, for frustrated magnetism in higher dimension, it is hard to find a similar composition. Therefore, in this work, we intend to begin with an alternative approach. We start from a Weyl semimetal system with eight Weyl cones, and couple our theory with a classical rotor. Each component of the rotor couples with the fermion bilinear and acts as a mass term. Once we integrating out the fermions, one obtains a Wess-Zumino-Witten theory for the fluctuating rotor\cite{xu2013wave,bi2014bridging,you2014symmetry}. The WZW term of the rotor originates from the special band structure of Weyl semimetal who contains monopole in momentum space around the Weyl points.

In the next two paragraph, we would investigate the phase transition theory between different symmetry breaking states in detail. The effective theory can be written in terms of LGW type theory plus an emergent gauge field coupling with the defect current. The emergent gauge field decorates the defect of an order parameter with a quantum number and the WZW term emerges when approaching the criticality. In addition, we can also consider an alternative case where we condense the double defects without carrying any quantum number. Such double defects condensation restore all symmetries and drive the theory into $Z_2$ topological order.

\subsection{Transition between SC and O(4) symmetry broken}
In this part, we would investigate a type of quantum phase transition beyond LGW paradigm. The transition connects a superconductivity state with an O(4) symmetry broken state\cite{moon2015competing,moon2012skyrmions,grover2008topological}. The superconductivity is realized by charge $2e$ SU(2) soliton condensation, contrary to the conventional BCS paired electron condensate. The SU(2) soliton can be view as the topological excitation of an O(4) rotor field in $3+1d$ which gives homotopy mapping $\pi_3(S^3)=Z$. The charged SU(2) soliton condensation disorders the O(4) rotor but meanwhile breaks charge U(1). In order to affix the U(1) charge to the soliton, we couple the O(4) rotor to the Weyl semimetals\cite{abanov2000theta}.

\begin{figure}[h!]
 \centering
  \includegraphics[width=0.2\textwidth]{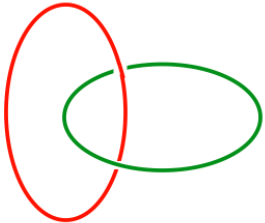}
 \caption{The linking between two vortex loops forms an SU(2) soliton. Here the red and green loops represents the gauge field $a$ and $b$ coming from two U(1) degree of freedom. }
 \label{link}
 \end{figure}

The Weyl semimetal we start with contains 8 Weyl points while 4 of them has left(right)-hand chirality.
These Weyl cones couple with an O(4) rotor $\vec{n}=(n_1,n_2,n_3,n_4)$ as,
\begin{align} 
&H=\Psi^{\dagger}_{\bm{k}}(\sigma_x \tau_z k_x+\sigma_y \tau_z k_y+\sigma_z \tau_z k_z+n_1\tau_x \nonumber\\
&+n_2\tau_y \pi_x \mu_y+n_3\tau_y \pi_y+n_4\tau_y \pi_z \mu_y)   \Psi_{\bm{k}}
\label{fermions}
\end{align}
$\sigma, \tau,\pi$ are Pauli matrices acting on the flavor index of the Weyl cones. When the rotor is in the ordered phase as $\vec{n} \neq 0$, the Weyl fermions are gapped. The ordered $\vec{n}$ vector acts as CDW order parameter which nests two Weyl cones with opposite chirality.
The O(4) rotor has a topological defect equivalent to an SU(2) soliton\cite{you2015topological,xu2013wave}. As one can write the O(4) degree of freedom in terms of the SU(2) matrix $U=n_4 I+i\sum^3_{i=1} n_a \sigma_a$, the SU(2)  soliton describes the O(3) skyrmion(for component $n_1,n_2,n_3$) living in the domain wall of $n_4$. As an alternative, one can also view the SU(2) soliton as a linking between two vortex loops by decomposing the O(4) rotor to $U(1)\times U(1)$ as Fig \ref{link}. By representing $a$ and $b$ as the gauge field for the $U(1)\times U(1)$ degree of freedom, the soliton density $J_{0}\sim\epsilon^{ijkl} \epsilon^{xyz}  n_i \partial_{x} n_j \partial_{y} n_k \partial_{z} n_l $ can be mapped to the linking number between two vortex loops in $3d$ space $\epsilon^{xyz} a_{x} \partial_{y} b_z$.

The proliferation of the SU(2) soliton disorder the O(4) rotor. However, as the rotor couples with the Weyl fermion, one needs to  carefully inspect the quantum degree of freedom carried by the soliton. Since the amplitude of the rotor is fixed, the fermion is always gapped even in the O(4) disorder phase. Coupling the fermion with the electromagnetic field and integrating out the fermion, we obtain the effective theory of the O(4) skyrmion as,
\begin{align} 
&\mathcal{L}=\frac{1}{g}|\partial_{\mu} n_{i}|^2+ 2 A_{\mu} J^{skyr}_{\mu}+i 2\pi H_3[\vec{n}]+...\nonumber\\
&J^{skyr}_{\lambda}=\frac{1}{12\pi^2}\epsilon^{\mu \nu \rho \lambda} \epsilon^{ijkl}   n_i \partial_{\mu} n_j \partial_{\nu} n_k \partial_{\rho} n_l 
\label{soliton}
\end{align}
$J^{skyr}$ is the SU(2) soliton current which couples with the electromagnetic field and carries charge $2e$. Before we condense the soliton to restore the O(4) rotation symmetry, we have to make sure the soliton here is a boson. $H_3[\vec{n}]$ in Eq.\eqref{soliton} is a topological invariant of the mapping from the spacetime into the target space $S^3$\cite{abanov2000theta}. There are only two homotopy classes $\pi_4 (S^3) = Z_2$. This geometric phase term indicates the soliton is a boson as its self rotation gives a phase of $2\pi$. 

Another way to investigate the statistics of the soliton is to decompose the O(4) vector into $O(3)\times Z_2=(n_1,n_2,n_3) \otimes n_4$. When there appears a domain wall of $n_4$, there is a $2d$ Dirac fermion theory localized near the domain wall,
\begin{align} 
&H=\Psi^{\dagger}_{\bm{k}}(\sigma'_x  k_x+\sigma'_z k_y \nonumber\\
&+n_1\sigma'_y \pi'_x \mu'_y+n_2\sigma'_y  \pi'_y+n_3\sigma'_y  \pi'_z \mu'_y)   \Psi_{\bm{k}}
\end{align}

 \begin{figure}[h!]
 \centering
  \includegraphics[width=0.4\textwidth]{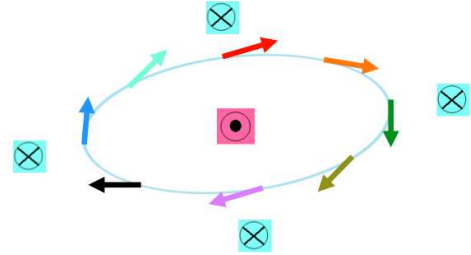}
 \caption{The skyrmion living on the domain wall plane.}
  \label{skyrmion}
 \end{figure}
 
Now we further decompose the O(3) vector into $O(2)\times Z_2=(n_1,n_2) \otimes n_3$.  The  O(3) skyrmion configuration can be illustrated as Fig \ref{skyrmion}, the domain wall for $n_3$ forms a loop and $(n_1,n_2)$
has a nonzero winding number along the domain wall loop. In the domain wall loop, the $1d$ fermion theory can be written as,
\begin{align} 
&H=\Psi^{\dagger}_{\bm{k}}(\tilde{\sigma}_x  k_x
+n_1\tilde{\sigma}_y \tilde{\pi}_z+n_2 \tilde{\sigma}_z \tilde{\pi}_z)   \Psi_{\bm{k}}
\end{align}
If $(n_1,n_2)$ does not have a winding number, the 1d fermion in the domain wall is fully gapped. Once we turn on a winding number of $2\pi$ along the domain wall loop corresponding to the creation of an SU(2) soliton for the O(4) rotor, the fermion spectrum would contain two zero modes which lead to the level crossings at $k=0$. Each additional fermion zero mode would change the fermion parity of the ground state. Consequently, the SU(2) soliton does not change the fermion parity and hence the soliton is bosonic. 

The SU(2) soliton in our theory can be mapped into a Hopf soliton. Consider an O(3) vector who transform as a spin 1 representation of the SU(2) group. The SU(2) gauge transformation for creating the SU(2) soliton can be mapped to the O(3) Hopf soliton.  Such mapping is illustrated as Fig \ref{link}. First, we decompose the O(4) rotor to $U(1)\times U(1)$. Each U(1) has a vortex loop and the linking between the two vortex loop is the SU(2) soliton. Now we bound the two vortex loop together as a ribbon, the linking now becomes the self-twist of the ribbon which is exactly the O(3) Hopf soliton\cite{you2015topological,ran2011fermionic}.

If we condensed the SU(2) soliton to restore the O(4) rotation symmetry, we would meanwhile break the charge U(1) and drive the theory into a superconducting state. Such superconductivity is induced by charge soliton condensation instead of the traditional fermion pairing scheme.

\subsection{Attacking from the superconductivity side}
In our previous discussion, we had shown that the charged SU(2) soliton condensation restores the O(4) rotation symmetry and meanwhile drives the system into charge 2e superfluid state. In this section, we would try to start with the opposite trend, we start from the superconductivity phase of the Weyl semimetal and restore the charge U(1) symmetry by vortex line condensation. The vortex line contains $1d$ gapless mode  of the O(4) rotor and we would show explicitly how does the vortex condensation process concur with O(4) symmetry breaking.

We have in total 4 Weyl pairs with opposite chirality and we turn on intra-cone s-wave pairing to gap out the semimetal. For each SC Weyl cone pairs, one can write them in the Nambu basis $\Psi^{\dagger}=\chi_1+i\chi_2$ and the Hamiltonian is,
\begin{align}
&H=\sum^4_{i=1} \chi^T_{i,\bm{k}}(i \partial_x \sigma^{103} +i \partial_y \sigma^{303} 
+i \partial_z \sigma^{223} \nonumber\\
&+ O_1 \sigma^{210}+O_2 \sigma^{230} ) \chi_{i,\bm{-k}}  \nonumber\\
&\Delta=O_1+i O_2
 \label{sc}
\end{align}
$\Delta$ is the pairing field and $\sigma^{abc}=\sigma_a\otimes \pi_b \otimes \tau_b$ with $\tau$($\pi$) acting on the chirality(Majorana) index.
For each SC Weyl pairs, the vortex loop of the pairing field $\Delta$ contains a $1d$ helical Majorana mode. As we have 4 Weyl pairs in total, the vortex loop carries 4 helical Majorana mode coupling with the O(4) vector $\vec{n}$ as,
\begin{align} 
&H= \chi^T_{\bm{k}}(i \partial_x \sigma^{100} +n_1 \sigma^{312}+n_2 \sigma^{320}+n_3 \sigma^{332}+n_4 \sigma^{200})   \chi_{\bm{-k}}
\end{align}   
Integrating out the fermion inside the vortex, we would obtain $1+1d$ O(4) WZW term at $k=1$, which represents the critical spin 1/2 Heisenberg chain akin to the $SU(2)_1$ conformal field theory.
\begin{align} 
&S=\int dx dt \frac{1}{g}(\partial_{\mu} n_i)^2+\int_{0}^{1} du \frac{ \epsilon^{ijkl} }{12\pi} n_i \partial_{x} n_j \partial_{y} n_k \partial_{u} n_l
\end{align}
Next we would demonstrate that the vortex condensation would concur with O(4) rotation symmetry breaking. Different from our previous approach, the order parameter defect here is a line object instead of a point particle. The vortex line carries gapless boson mode(of the O(4) rotor) and it is hard to write down a straightforward theory describing the loop condensation in the presence of gapless modes. To detour this problem we borrow a domain wall from the O(4) rotor.

\begin{figure}[h!]
 \centering
  \includegraphics[width=0.2\textwidth]{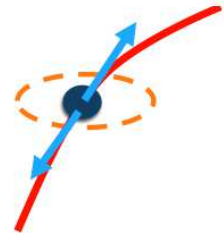}
 \caption{The domain wall living on the SC vortex line forms a `monopole defect'. The blue arrow is the domain wall while the red line is the vortex loop}
  \label{vortexdomain}
 \end{figure}

We first add an anisotropic term to break the O(4) rotor into $O(3) \times Z_2$. This can be achieved by adding a large mass term like $\alpha n_4^2$ so the $n_4$ component is suppressed to zero.  Once $n_4$ is disordered, we can assume the ground state is saturated with all superpositions of domain walls for the scalar field $n_4$. 
When the domain wall plane is perpendicular to the superconducting vortex line, the $Z_2$ domain wall and the U(1) SC vortex together forms a monopole defect as Fig \ref{vortexdomain}. 
Inside the monopole, there contains $0+1d$ O(3) WZW term at $k=1$, 
\begin{align} 
&S=\frac{1}{4}\int_{0}^{1} du \int dt~ \epsilon^{ijk}  n_i \partial_{t} n_j \partial_{u} n_k
\end{align}
This Wess-Zumino-Witten model at $k=1$ represents a spin $\frac{1}{2}$ degree of freedom for  $\vec{N}=(n_1,n_2,n_3)$ \cite{xu2013nonperturbative}.  Once we condense the bound state between $n_4$ domain wall and SC vortex as a `rough' monopole, the spinon inside the monopole would be condensed as well.
This restores the charge U(1) but meanwhile breaks the O(3) of $(n_1,n_2,n_3)$.

To describe the theory of such `rough' monopole condensation, we first write the spinon of $\vec{N}$ in $CP^1$ form.
\begin{align} 
&n_i=\frac{1}{2}z^* \sigma_i z
\end{align}
$z$ is a two component complex field with $|z_1|^2+|z_2|^2=1$. The $z$ field has a U(1) gauge symmetry $z \rightarrow e^{i\theta} z$. Thereby when writing the NL$\sigma$M of the O(3) rotor in terms of the $CP^1$ form, the spinon field $z$ automatically couples to a U(1) gauge field $a_{\mu}=z_i^* \partial_{\mu}z_i$. 

As the composite monopole(as a bound state of SC vortex and $n_4$ domain wall) is decorated with a spinon degree of freedom, one could assume the composite monopole field 
 carries a fundamental representation of SU(2)\cite{xu2012unconventional} and couples with the gauge field $a$. As a result, one can precisely use the $CP^1$ field theory to describe the monopole condensation transition,
\begin{align} 
&\mathcal{L}=|(\partial_{\mu}+ia_{\mu})z_{i}|^2+\mu |z_i|^2+\beta (\sum_i |z_i|^2)^2+\kappa f_{\mu \nu}f^{\mu \nu}
\label{aaa}
\end{align}
The $CP^1$  field represents the composite monopole degree of freedom.
$f_{\mu \nu}$ is the electromagnetic tensor of $a$. Here we have soften the constrain $|z_1|^2+|z_2|^2=1$ and replace it with the interaction $\beta$ term. When $\mu$ is positive, the spinon acquires a nonzero expectation value so the O(3) rotation symmetry is broken.
The SC to O(4) symmetry breaking transition is characterized by the $CP^1$ spinon condensate due to the fact that the `rough' monopole consist of SC vortex and $n_4$ domain wall carries a spinon of $\vec{N}=(n_1,n_2,n_3)$. The monopole condensation higgsed the gauge field of the spinon and the O(3) rotation symmetry is broken(which also breaks the original O(4)). As a result, the SC phase and the O(4) symmetry broken phase can be connected through a second order transition.

As the monopole composed from $n_4$ domain wall and SC vortex
carries a spinon of $\vec{N}=(n_1,n_2,n_3)$, the coupling between them can be written as,
\begin{align} 
&\mathcal{S} \sim   \int d^3 x dt ~a_{\lambda} J^{N}_{\lambda},\nonumber\\
&J^{N}_{\lambda} \sim \frac{1}{8\pi}\epsilon^{\lambda \rho \mu \nu }\epsilon^{jkl }\partial_{\rho}n_j^e\partial_{\mu}n_k^e\partial_{\nu}n_k^e, 
\end{align}
$J^{N}$ is the monopole current of $\vec{N}=(n_1,n_2,n_3)$ and $a$ is the U(1) gauge field for $z$. This coupling indicates there could appear an emergent Wess-Zumino-Witten term at the phase transition point.

At the critical region when both O(4) and SC are disordered, the SC order parameter together with the O(4) rotor has an emergent O(6) symmetry.
We can write the Weyl semimetal in the Nambu basis,
\begin{widetext}
\begin{align} 
&H= \chi^T_{\bm{k}}(i \partial_x \sigma^{10300} +i \partial_y \sigma^{30300}+i \partial_z \sigma^{22300}+ O_1 \sigma^{21000} 
+O_2 \sigma^{23000} 
+n_1 \sigma^{02110}+n_2 \sigma^{02130}+n_3 \sigma^{02122}+n_4 \sigma^{02202}) \chi_{\bm{-k}} \nonumber\\ 
\end{align} 
\end{widetext}
After we integrating out the fermion, criticality theory connecting two symmetry breaking phase is controlled by the O(6) WZW theory,
\begin{align} 
&S=\frac{1}{4} \int d^3 x dt~ [\frac{1}{g}|\partial_{\mu} n_{i}|^2\nonumber\\
&+ \int_{0}^{1} du \frac{2\pi}{\Omega^5} \epsilon^{ijklmn}  n_i \partial_{x} n_j \partial_{y} n_k \partial_{z} n_l \partial_{t} n_m \partial_{ u} n_n]
\label{wzw}
\end{align}


\subsection{Double vortex condensation, topological order}
In our last section, we had shown that in the SC phase, the condensation of vortex loop restores the charge U(1) but meanwhile breaks the O(3) symmetry. Then one may ask it is possible to restore the charge U(1) without breaking O(3)\cite{balents1999dual}?
The answer is Yes. Instead of condensing single vortex, we bound two vortices together and condense the double vortex lines. The double vortex line carry two copies of the critical boson chain described by a NL$\sigma$M  with $\Theta=\pi$\cite{you2016stripe}. By turning on ferromagnetic/antiferromagnetic interaction between the two O(3) rotors, we can gap the two critical boson chain into a rotation invariant state. They are two ways to gap out the critical chain, which finally drives the system into a O(3) NL$\sigma$M with either $\Theta=0$ or $\Theta=2\pi$.  Both of these two gapped chains are rotation invariant, so the double vortex line condensation does not break the rotation symmetry. In addition, the effective theory exhibit $Z_2$ topological order which can be described by the BF theory,
\begin{align} 
&\mathcal{L}=\frac{1}{4\pi} \epsilon^{\mu \nu \rho \lambda} B^{\mu\nu} \partial_{\rho} a_{\lambda}
\end{align}
Here $\epsilon^{\mu \nu \rho \lambda} \partial_{\rho} a_{\lambda}$ is the current for a single SC vortex and $\epsilon^{\mu \nu \rho \lambda}  \partial_{\rho} B^{\mu\nu}$ is the O(3) monopole current. This term indicates the monopole and vortex loop has mutual semion statistics. 

There are two ways to gap out the gapless mode inside the double vortex line and they end up with the same 3d topological order after double vortex condensed. However, these two states can be distinguished on the boundary.
The end point of vortex line on the boundary is a vortex particle. If the bound state of two vortex lines are gapped into 
$\Theta=2\pi$ phase, the end point of double vortex on the boundary carries spin 1/2 degree of freedom as a Kramers doublet.  Thus, the time reversal operator acting on the double vortex line in a projective way. The surface theory thereby contains either gapless mode or anomalous topological order. This $Z_2$ topological ordered phase is associate with the $3d$ $Z_2 \times \mathcal{T}$ invariant boson SPT phase after $Z_2$ symmetry gauging.

\section{The PDW superconductor melting, 3d string-membrane condensation}
In our previous discussion, we investigate the transition from SC to O(4) rotation symmetry broken states by vortex line condensate. In this part, we assume the superconducting state from Weyl cone pairing is nonuniform in space as the Cooper pairs condense with a global momentum\cite{berg2009charge}. As a result, the consequent SC order parameter has a pair density wave(PDW) structure whose pairing amplitude modulates in space. 

Assume the SC order parameter modulates along $z$ direction. One can write the pairing of the Weyl cones in the Nambu basis as\cite{you2016stripe},
\begin{align} 
&H=\chi^T_{i,\bm{k}} (i \partial_x \sigma^{10300} +i \partial_y \sigma^{30300}+i \partial_z \sigma^{22300} +O_1 \sigma^{21000} \nonumber\\
&+n_1 \sigma^{02212} +n_2 \sigma^{02232} +n_3 \sigma^{02220} +n_4 \sigma^{02100}  ) \chi_{i,\bm{-k}}  \nonumber\\
&\Delta=O_1=|O_1| \cos (qz)
\end{align}
Here we choose a specific gauge when the PDW order is real and therefore $O_2=\langle \chi^T_{i,\bm{k}}\sigma^{23000}\chi_{i,\bm{-k}}  \rangle =0$.

This PDW state has a slab configuration whose amplitude modulates along the PDW wave vector. In the nodal plane of the PDW state at $qz=(n+1/2)\pi$, the SC amplitude is zero and the Majorana fermions merely couple with the O(4) rotor.
We first break the O(4) vector down to O(3) by developing a nonzero expectation value of $n_4$. In addition, we assume $n_4=m$ is positive and large compared to the fluctuating O(3) rotor.

Thereupon, the fermions are gapped inside the nodal plane so one can integrate out the fermion band to obtain the effective theory of the O(3) rotor in the $2d$ nodal plane. The disordered O(3) rotor is actually in the SPT phase whose effective theory is a descendent of the topological NL$\sigma$M in $2d$. 

To demonstrate this, first we look at the fermions inside the nodal plane. The nodal plane is a domain wall of $O_1$. There exists 8 copies of $2d$ Majorana cones localized inside the nodal plane who couple with the rotor as,
\begin{align} 
\label{zljkh}
&H= \chi^T_{\bm{k}}(i \partial_x \sigma^{1000} +i \partial_y \sigma^{3000}+m\sigma^{2300}+O_2 \sigma^{2100}\nonumber\\
&+n_1 \sigma^{2212}+n_2 \sigma^{2220}+n_3 \sigma^{2232})   \chi_{\bm{-k}} 
\end{align}
Here $m$ is the expectation value of $n_4$ which is already polarized.

Integrating out the fermions in the nodal plane, one obtains a topological NL$\sigma$M with $\Theta=2\pi$ whose ground state wave function can be written in terms of  all coherent configurations of the O(4) degree of freedom $(n_1,n_2,n_4,O_2)$. The coefficient of each configuration is the O(4) WZW term. 

However, inside the nodal line, the SC amplitude is zero so $O_2=0$. We can apply this constraint by adding a large mass term $|u|O^2_2$ to push $O_2$ to zero. Then the effective theory is merely the dynamics of the O(3) rotor. 
The wave function of the O(3) rotor in the nodal plane can be written in terms of the coherent sum of all rotor superpositions and the coefficient for each coherent state counts whether there are even or odd number of skyrmions in each configuration. 
 \begin{align} 
 \label{sign}
&|GS \rangle= \int D[\vec{n}] e^{i\frac{2\pi}{\Omega^{2}} \int d x^2   \epsilon^{ijk} n_i \partial_{x}n_j \partial_{y} n_k} |\vec{n} \rangle \nonumber\\
&|\vec{n} \rangle=(n_1,n_2,n_3)\nonumber\\
&Z_2: (n_1,n_2,n_3,O_2)\rightarrow -(n_1,n_2,n_3,O_2)
\end{align}

 Such wave function illustrates an SPT state protected by $Z_2$ symmetry where the sign structure of the coherent sum in Eq. \eqref{sign} can never be erased out as long as the symmetry is unbroken\cite{levin2012braiding,xu2013wave}. The nodal plane is therefore decorated with a topological paramagnetic state.

Now we are about to disorder PDW state and condense the nodal plane to restore the spatial symmetry.
 The melting procedure can be realized by dislocation and disclination proliferation, which bends the nodal plane into arbitrary close membrane configuration. The condensation of nodal membrane restores the rotation/translation symmetry. We assume there is some thermal or quantum fluctuation which effectively generates positive interaction between disclination(dislocation) and therefore triggers a tendency to condense them. During the condensation, the origin Goldstone mode($\phi$) associate with the translation symmetry broken is gapped by the vortex tunneling term cos(n$\phi$). The coherence of the condensed dislocations and disclinations generates a coherent state of all types of nodal membrane configurations. Meanwhile, since the nodal membrane separates the positive and negative pairing amplitude, the nodal line must be closed in the bulk in the GS, otherwise there must be a half vortex of the pairing field associated with it. We focus on the situation where the GS is vortex free so all the nodal line must form a close-loop configuration.

 After nodal membrane condensed, the GS is the superposition of all close membranes decorated with a 2d SPT state(topological paramagnetic). One can write the wave function in terms of all coherent superposition of O(4) soliton configurations where the coefficient of each configuration carries a sign structure counting the total parity of solitons.
 
 \begin{align} 
&|GS \rangle= \int D[\vec{n}] e^{i\frac{2\pi}{\Omega^{3}} \int d x^3   \epsilon^{ijkl} \tilde{n}_i \partial_{x} \tilde{n}_j \partial_{y} \tilde{n}_k\partial_{z} \tilde{n}_l} |\vec{\tilde{n}} \rangle \nonumber\\
& |\vec{\tilde{n}} \rangle=(n_1,n_2,n_3,O_1)
\end{align}
This wave function can be regarded as a consequence of both loop condensate and membrane condensate in $3d$. If we only look at $O_1$, the GS wave function is the condensation of all close nodal membrane of $O_1$, where the membrane itself is decorated with an O(3) topological paramagnetic state. Alternatively, if we focus on the skyrmion flux $B_k \sim n_a \partial_{i} n_b \partial_{j} n_c$ degree of freedom, the GS wave function is the condensation of all close flux loops where the loop contains fluctuating domain walls of $O_1$\cite{moradi2015universal}. This wave function is akin to the $3d$ boson toric code model\cite{moradi2015universal} whose wave function can be written in terms of either membrane or loop condensate. Here and after, we would argue that there is $Z_2$ topological order in this melted PDW state.

Before we melt the PDW, there are three types of deconfined excitations, a $2\pi$ dislocation, a $2\pi$ SC vortex and a $\pi$ dislocation $+$ $\pi$ SC vortex bound state.  The $\pi$ dislocation and $\pi$ SC vortex bound state is associated with the condition where the nodal plane has an open boundary line bounded with a $\pi$ SC vortex line. The half-vortex traps an electromagnetic flux $\pi/2$. After we melt the PDW slab, the $\pi$ dislocation $+$ $\pi$ SC vortex bound state still remains as a deconfined excitation. Thus, one can conclude that the open membrane of $O_1$ contains a loop excitation carrying electromagnetic flux  $\pi/2$.

Beyond such deconfined loop excitation as the boundary of opened membrane, we can also find a deconfined particle excitation as the end point of an open string. If we write the O(3) rotor degree of freedom in terms of the $CP^1$ representation, 
\begin{align} 
&n_i=\frac{1}{2}z^{*T}\sigma_i z\nonumber\\
&a_{\mu}=z_i^*\partial_{\mu} z_i,\nonumber\\
& B_k=\epsilon^{ij} \epsilon^{abc} \frac{1}{8\pi}n_a \partial_i n_b \partial_j n_c= \frac{1}{2\pi} \epsilon^{ijk} \partial_i a_{j}
\end{align}
The skyrmion flux can be written in terms of the electromagnetic flux of an U(1) gauge field $a$.
In the PDW melted phase, the flux line condenses and the monopole excitation is deconfined. The monopole of the gauge field $a$ is associated with the hedgehog configuration of the O(3) rotor. Here and after, we would demonstrate that the monopole at the end of skyrmion flux and the loop on the boundary of membrane has mutual semion statistics.

To proceed, we would prove that the O(3) monopole is a dyon which carries electric charge $2e$.
Starting from the fermion model we worked on, the O(3) rotor couples with the fermion as,
\begin{widetext}
\begin{align} 
&H=\Psi^{\dagger}_{\bm{k}}(\sigma_x \tau_z k_x+\sigma_y \tau_z k_y+\sigma_z \tau_z k_z+m\tau_x
+n_1\tau_y \pi_x \mu_y+n_2\tau_y \pi_y+n_3\tau_y \pi_z \mu_y)   \Psi_{\bm{k}}
\end{align}
\end{widetext}
When the mass term is positive and much larger than O(3) vector, after integrating out the fermions, the effective theory between O(3) monopole and the electromagnetic field is expressed in terms of the axion electrodynamics,\cite{qi2008topological},
\begin{align} 
\frac{4\Theta}{4\pi^2} dA da,(\Theta=\pi),  
\end{align}
We have a factor of $4$ since the Hilbert space is quadrupled compared to those of 3D TI. The axion term indicates when we have a monopole of O(3), there traps a polarized charge $2e$ inside the monopole\cite{qi2008topological}.

In the GS, the flux line of $a$, interpreted as the skyrmion flux, can be written as a coherent sum for all close flux loop configurations so the monopole as the end point of the open flux string is a deconfined particle excitation. 
The monopole carries charge 2e, while the boundary loop of the open membrane trap EM flux $\pi/2$. The winding between the monopole and the loop accumulates a $\pi$ Berry phase. This demonstrates that the melted PDW SC is a boson toric code theory in $3d$\cite{wen1990topological}. 

\begin{figure}[h!]
 \centering
  \includegraphics[width=0.4\textwidth]{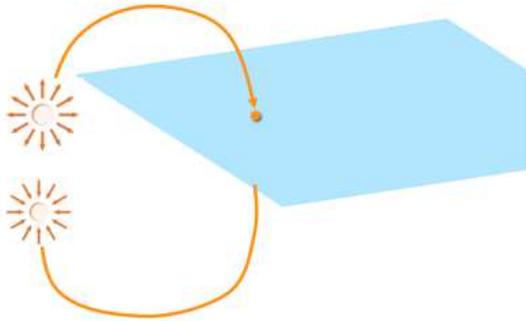}
 \caption{We create a pair of monopole/anti-monopole on the left, take one monopole braiding around the membrane boundary line and annihilate the pair. The blue plane is the open nodal membrane. The orange line is the trajectory of the monopole. Once the monopole goes across the nodal membrane, it adds additional skyrmion number on the plane so the wave function of the topological paramagnetic acquires a minus sign.}
  \label{line}
 \end{figure}

Another way to confirm the $\pi$ statistics between flux loop and monopole is illustrated as Fig \ref{line}. Imagine we create a pair of monopole/anti-monopole, take one monopole winding around the boundary line of the nodal membrane and finally annihilate the monopole/anti-monopole pair. When the monopole goes across the nodal membrane, it creates a skyrmion on the nodal plane. As is demonstrated in the wave function at Eq. \eqref{sign}, the topological paramagnetic state on the nodal plane can be written as the coherent sum of all skyrmion configurations while each additional skyrmion contributes a minus sign factor. Hence, the braiding procedure creates addition skyrmion on the nodal plane accommodates with a global minus sign factor in the wave function. This indicates  flux loop and monopole has mutual $\pi$ statistics.

In conclusion, the PDW melting transition connects a symmetry breaking state with a $3d$ topological ordered state. The essence of such transition lies in the fact that the nodal plane is decorated with a topological paramagnetic state, and nodal membrane condensation give rise to a boson SPT protected by $Z_2\times \mathcal{T}$ symmetry, whose effect theory is described by the O(5) NL$\sigma$M\cite{bi2015classification,xu2013wave,you2016stripe},
\begin{align} 
\label{3dnlsm}
&\mathcal{L}= \sum^5_{a=1}\frac{1}{g}(\partial_{\mu} N_{a})^2+\frac{i2\pi}{\Omega^4} \epsilon^{ijklm} N^i \partial_{x}  N^j \partial_{y} N^k \partial_{z}  N^l  \partial_{t} N^m \nonumber\\
&\vec{N}=(O_1,O_2,n_1,n_2,n_3)                  \nonumber\\
&\mathcal{T}: n_{1,2,3} \rightarrow  n_{1,2,3}; ~Z_2:  (O_1,O_2) \rightarrow  -(O_1,O_2)
\end{align} 
If we gauge the $Z_2$ symmetry, the half SC vortex of $(O_1,O_2)$ couples with the $Z_2$ gauge flux. This flux loop is associate with the open membrane boundary of $O_1$, which
exhibit nontrivial loop-particle braiding with the monopole. Consequently, the $Z_2$ topological order in our theory is related to a $3d$ SPT state after symmetry gauging. The half vortex $+$ half dislocation pair plays the role as the bound state between $Z_2$ gauge flux and half SC vortex, and the PDW state we start with is crucial as it provides a system with deconfined half SC vortex excitation\cite{berg2009charge}. The PDW melted state is thereby equivalent to a gauged $3d$ boson SPT phase.

\subsection{Three loop statistics in this model}
If we do not fix $n_4$ in equation Eq. \eqref{zljkh}, the O(4) rotor coupling with the Majorana cones acquires an O(4) $\Theta$ term at $\Theta=\pi$. Breaking the O(4) symmetry to $U(1) \times U(1)$, the theory is gapped with mutual semion topological order. The nodal plane of the PDW is therefore decorated with the $Z_2$ topological order. After we melt the PDW slab, there are three types of deconfined excitations: The vortex line for $n_1,n_2$(label as $a$), the vortex line for $n_3,n_4$(label as $b$), and the half-vortex half dislocation of the SC(label as $c$). When the braiding between vortex lines of $a$ and $b$ are penetrated by the half vortex of the SC, the 3 loop braiding acquires a $\pi$ Berry phase, as a signature of the nontrivial 3 loop statistics. The trajectory of such 3-loop statistics, can also be expressed in terms of the particle loop statistics where the `particle object' is composed by the two vortex loop $a$ and $b$ linking to each other. Else, if we project the two vortex loop into a global U(1) degree of freedom, the linking of the vortex loops becomes the self-twist of a ribbon band. The 3-loop statistics then reduce to the particle-loop statistics between a self-twist ribbon and SC half vortex\cite{wang2014braiding}.

\section{Ising order percolation}

\subsection{Saturation of surface topological order to the bulk}
In this section, we focus on the phase transition between SPT to topological order phases. The relation between SPT phases and topological order states is well explored in terms of symmetry gauging\cite{you2015topological,bi2014anyon,levin2012braiding}, where the gauge flux exhibit nontrivial statistics. In these precedents, the transition between SPT and topological order states could be characterized as the confinement-deconfinement transition of the gauge flux. 

At this point, we would start with another approach to bridge the relation between SPT and topological phase in $3d$. We start with a certain type of $3d$ SPT state whose surface can exhibit topological order. By saturating the $2d$ domain wall between SPT and trivial phase into the bulk, the system is soaked with domain wall membranes decorated with $2d$ topological order. Such domain wall condensate drives the SPT state into a topological ordered phase with nontrivial loop-particle statistics. 

To study such phase transition explicitly, we begin with the microscopic model studied in Eq.\eqref{fermions}.
\begin{align} 
&H=\Psi^{\dagger}_{\bm{k}}(\sigma_x \tau_z k_x+\sigma_y \tau_z k_y+\sigma_z \tau_z k_z+m\tau_x \nonumber\\
&+n_1\tau_y \pi_x \mu_y+n_2\tau_y \pi_y+n_3\tau_y \pi_z \mu_y)   \Psi_{\bm{k}}
\end{align}
Assume $m$ is positive and its amplitude is large compared to the O(3) rotor $(n_1,n_2,n_3)$. When the O(3) rotor is disordered, the theory describes a time reversal invariant topological insulator inherited from the chiral symmetry breaking of the Weyl fermions. The surface state of this topological insulator, who contains 4 Dirac cones, can be regarded as the domain wall of $m$ where the mass changes the sign. If we turn on the s-wave superconductivity of the surface Dirac cone, the surface state is thereby gapped and the SC order $\Delta$ breaks $\mathcal{T}$ and charge U(1) symmetry. The SC surface can be written in the Nambu basis as,
\begin{align} 
&H= \chi^T_{\bm{k}}(i \partial_x \sigma^{1000} +i \partial_y \sigma^{3000}+ O_1 \sigma^{2100} \nonumber\\ 
&+O_2 \sigma^{2300} 
+n_1 \sigma^{2212}+n_2 \sigma^{2232}+n_3 \sigma^{2220}) \chi_{\bm{-k}} \nonumber\\ 
& \Delta=O_1+iO_2,~\mathcal{T}: \chi \rightarrow \mathcal{K}i\sigma^{2000}\chi
\end{align} 
The nonzero surface SC order $O_1,O_2$ breaks charge U(1) and $\mathcal{T}$. To restored the broken symmetry, one has to condense the vortex of the superconductivity order. However, the vortex of the SC order carries a spinon degree of freedom of the O(3) rotor. This can be easily verified by the O(5) WZW term at $k=1$ after we integrate out the fermion. The $O(5)_1$ WZW term, composed from the SC order together with an O(3) rotor, indicates the SC vortex contains an $O(3)_1$ WZW term at $0d$, which is exactly the spinon of the rotor. The spinon decorated vortex is a Kramers doublet whose condensation breaks the $\mathcal{T}$ symmetry. Thus, to restore the charge U(1) and $\mathcal{T}$, one needs to condense double SC vortex.  After double vortex condensate, the single vortex is a deconfined excitation carrying Kramers doublet. Meanwhile, the vortex between $n_1,n_2$ also carries a Kramers doublet as it contains an $O(3)_1$ WZW term of $n_3,O_1,O_2$. The vortex of $n_1,n_2$ has mutual semion statistics with the SC vortex. If we add a mass term for $n_3$  which suppress the $n_3$ component to zero, the $O(5)_1$ WZW term is therefore reduced to the O(4) topological $\Theta$ term at $\Theta=\pi$ with $U(1)\times U(1)$ anisotropy. The $\Theta$ term denotes the mutual semion statistics between the vortex of $n_1,n_2$ and SC vortex, both of which carries Kramers doublet. This surface topological order is known as the $[e\mathcal{T}m\mathcal{T}]$ surface state\cite{metlitski2014interaction,wang2014interacting}.

To drive SPT state into a topological ordered state, we proliferate domain wall of the mass term $m$ in the bulk. Such domain wall separates SPT states with vacuum and therefore supports the $[e\mathcal{T}m\mathcal{T}]$  $Z_2$ topological order we discussed in the previous content. The condensation of 
$[e\mathcal{T}m\mathcal{T}]$ state decorated domain wall drives the theory into a 3d topological order with nontrivial loop-particle braiding.
To demonstrate this, we first write down the effective theory after domain wall condensate. As the domain wall contains $Z_2$ topological order in $2d$, one can write down an O(5) $\Theta$ term in $3d$ with $U(1)\times U(1) \times Z_2$ anisotropy to characterize their braiding,
\begin{align} 
&\mathcal{L}_{\Theta} \sim \frac{ \epsilon^{ijklh} }{\Omega^4} n'_i \partial_{x} n'_j \partial_{y} n'_k \partial_{z} n'_l  \partial_{ t} n'_h \nonumber\\ 
&\vec{n'}=(n_1,n_2,O_1,O_2,m)
\label{scvortex}
\end{align}
The intersection between vortex line of $n_1,n_2$ and domain wall of $m$ at the perpendicular direction forms a particle object, we can thereby define the gauge field associated with the particle and loop current.
\begin{align} 
&B_{\mu \nu} \sim n_1 \partial_{\mu} n_2 \partial_{\nu} m\nonumber\\ 
&a_{\mu}\sim \epsilon^{ij} O_i\partial_{\mu} O_j
\end{align}
The O(5) $\Theta$ term then becomes the BF term $\frac{1}{4\pi}B\wedge da$. The monopole formed by the vortex and the domain wall is a deconfined particle excitation which has mutual semion statistics with the SC vortex.

At this point, we had demonstrated that the transition between SPT to topological ordered phase could be realized by domain wall condensation where the surface topological order goes into the bulk and saturates. To go backward, we start from the topological ordered state and pull it back to the SPT(or trivial phase) by anyon condensate. 

In order to confine the monopole and trivialize the topological order in the bulk, one could condense the SC vortex loop. However, as the SC vortex loop carries gapless mode, the condensation would give rise to a gapless photon phase. In order to obtain a trivial gapped phase, one has to get rid of the gapless mode inside the vortex loop.

As is demonstrated by Eq. \eqref{scvortex}, the SC vortex and the monopole(formed by the domain wall and vortex of $n_1,n_2$) together forms an O(5) $\Theta$ term at $\Theta=\pi$. If we develop a SC vortex line along the $i$ direction, the vortex line contains an O(3) NL$\sigma$M with a $\Theta$ term at $\Theta=\pi$. This exactly describes the spin 1/2 chain with short-ranged interaction whose ground state is either gapless or symmetry breaking. The O(3) degree of freedom composed by $(n_1,n_2,m)$ could be written in terms of the $CP^1$ field characterizing the spinon degree of freedom. If spinon does not condense, the SC vortex carries $1d$ gapless mode akin to the spin 1/2 AF chain. To get rid of the gapless mode, one needs to condense the spinon $z$ which breaks the O(3) rotation symmetry.  After the spinon condense, the vortex line no longer carries gapless mode. Without loss of generality, we assume the spinon is condensed and $m$ is nonzero.  At this stage, the condensation of the SC vortex line gives rise to either SPT or trivial phase, depending on the sign of $m$. As long as $m$ is ordered, the domain wall membrane is confined in the bulk. When $m$ is positive, the theory describes an SPT phase protected by $\mathcal{T}$ symmetry while negative $m$ is associated with the trivial phase. The interface between the trivial-SPT phase is the domain wall of $m$ which supports $[e\mathcal{T}m\mathcal{T}]$ surface topological order.
When $m$ is positive, the wave function after SC vortex condensate can be written as\cite{xu2013wave},
\begin{align} 
|GS \rangle= \int D[\vec{n}] e^{i\frac{2\pi}{\Omega^{3}} \int d x^3   \epsilon^{ijkl} n_i \partial_{x}n_j \partial_{y} n_k\partial_{z}n_l} |\vec{n} \rangle
\end{align}
One can combine the vortex of SC together with the $n_1,n_2$ degree of freedom and express them in terms of SU(2) matrix $U=n_1I +i(n_2 \sigma_x +O_1\sigma_y+O_2\sigma_z) $. The wave function is  a coherent sum of all SU(2) soliton configurations and the sign factor for each configuration counts the number of solitons. Each additional soliton in the configuration gives a minus sign to the factor\cite{xu2013wave}.
In addition, we can also regard the wave function as a superposition of SC and $n_1,n_2$ vortex loop saturating in the bulk. The coefficient of each superposition counts the linking number between the SC and $n_1,n_2$ vortex loop while each linking contributes a minus sign.

As a summary, we investigate the transition from SPT to topological order phase in $3d$ via domain wall condensation from 4 copies of $3d$ topological insulators. The domain wall, acts as the interface between SPT and trivial phase of TI supports $[e\mathcal{T}m\mathcal{T}]$ surface topological order. As long as the domain wall goes into the bulk and saturates, the bulk exhibit topological order with nontrivial loop-particle statistics. On the opposite side, if we start from the $3d$ topological phase, one can confine the monopole by loop condensate.  As the vortex loop carries gapless mode, to obtain a gapped trivial phase, one have to condense the charge(spinon) associate with the monopole to get rid of the gapless mode inside the vortex.  The correspondence phase after vortex loop condensate is either a trivial or SPT phase depending on the symmetry breaking order after spinon condensate.

\subsection{Transition from Topological order to SPT in 3d, a holographic view}
In our previous discussion, we introduce a phase transition between SPT and topological order in $3d$  boson topological insulators via surface topological order proliferation. This idea could be extended to other type of SPT phases. 

Imagine we have a $3d$ SPT state whose surface exhibit topological order described by the K-Matrix $\frac{K_{ij}}{4\pi} a^i\wedge d a^j$(This K-Matrix only characterize the topological order of the quasiparticles without assigning any symmetry). If we proliferate the domain wall between SPT to vacuum into the bulk, the corresponding topological order decorated domain wall condensation drives the theory into a $3d$ topological ordered phase. The particle-loop braiding in the bulk can be written in terms of the BF term $\frac{K_{ij}}{4\pi} a^i\wedge d B^j$\cite{jian2014layer}. $B$ is two form gauge field and $ d B^j$ is a monopole object composed of the vortex line of $a_j$ with the domain wall intersecting on the vortex line.  

On the opposite trend, if we start the from the topological order side and condense the vortex loop $a^i$ to confined the monopole and trivialize the phase, we have to confine the domain wall as well otherwise the theory is either symmetry broken or gapless.  The argument works as follow. As the domain wall condense in the bulk, the vortex line of $a_j$ is saturated with the domain wall. At each intersection between domain wall and vortex , one can regard it as the vortex core of $a_j$ at the surface of the SPT phase. Such vortex core must carry some nontrivial quantum number of the symmetry who protects the SPT state, otherwise one could condense the anyon on the surface to obtain a gapped symmetric trivial surface  which is forbidden in SPT system. Thus, one cannot simply condense the vortex line as it would either drive the theory into gapless state or breaks the symmetry who protects the SPT phase. As a result, one has to suppress the domain wall in the bulk before vortex loop condensate. When the domain wall is suppressed, the bulk goes back into the SPT or trivial phase. 
 
 The transition between SPT and topological order via surface topological order saturation can be generalized to a class of SPT phase based on our previous argument. However, the above argument only applies when the SPT state surface does not have perturbative anomaly. If the surface has perturbative anomaly, the surface state cannot exhibit symmetry invariant topological order as the anomaly matching condition tells us the perturbative anomaly is always associate with gapless mode in IR. 
 
In addition, the transition between SPT and topological ordered state in $3d$ can be mapped into a symmetry breaking transition in $4d$\cite{burnell2012phase,tsui2015topological,tsui2015quantum,chen2013critical}. Imagine we have a $Z^T_2$ variable `$m$' in $4d$ space. The theory is in the disordered phase so the domain wall of $m$ proliferates in space. We now decorate the domain wall with a $3d$ SPT state protected by symmetry $G$(assume $G$ is an internal symmetry). After the domain wall condensate, the theory is a $4d$ SPT protected by the $G \times Z^T_2$ symmetry. This is exactly the idea of the decorated domain wall construction of SPT states\cite{chen2014symmetry,tsui2015quantum,tsui2015topological}. On the boundary of the $4d$ SPT, if the $3d$ SPT state protected by symmetry $G$ supports surface topological order in $2d$, the $3d$ boundary can be expressed as the domain wall condensation embellished with a $2d$ topological order. Consequently, the $3d$ boundary contains nontrivial loop-particle statistics. This is exactly the $3d$ topological order phase we studied in our previous content.
 
  If the $Z^T_2$ symmetry is broken, the domain wall percolation shrinks in $4d$ and the $m$ variable is ordered. The surface of such $Z^T_2$ ordered  bulk is either in the SPT or trivial phase with respect to the symmetry $G$. The argument flows as follow, if we have an interface between different Ising ordered phase of $m$, the interface as a $3d$ domain wall is an SPT phase  protected by symmetry $G$. The surface of such SPT contains topological order with obstruction where the topological quasiparticle carries nontrivial quantum numbers. As a result, the domain wall between the corresponding surfaces of the two Ising order phases contain anomalous topological order. Thus the two surface state must belong to different phase of matter as long as the symmetry $G$ is preserved.

To conclude, the topological to SPT phase transition in $3d$ can be mapped to the symmetry breaking of a scalar field in $4d$, provided that the domain wall defect of the scalar variable is decorated with the $3d$ SPT. However, such decorated domain wall scheme only works when the lower dimension SPT inside the domain wall follows the NDST\cite{tsui2015quantum}, otherwise the surface state can always be gapped without symmetry breaking.

\section{Conclusion and outlook}

In this paper, we elaborate several unconventional quantum phase transitions in Weyl semimetals coupling with fluctuating bosonic fields, whose criticality contains a WZW term or $\Theta$ term. The existence of such topological term in the critical region evidences the fact that a defect of the boson order parameter carries either a quantum number or a topological term. Consequently, once we start with a symmetry breaking state and proliferate the order parameter defect, the defect condensation would restore the original symmetry but meanwhile breaks another symmetry(if defect carries quantum number) or drives the state into $3d$ topological order(if defect carries topological term). 

The spirit of such exotic quantum phase transition
is the defect decoration. The emergent WZW term(or topological $\Theta$ term) appearing at the criticality is encoded in the decorated defect degree of freedom. Further, we also look into the phase transition between a $3d$ topological order to a trivial(or SPT) phase by proliferating the domain wall between SPT/trivial interface. As long as the SPT surface states support topological order, the domain wall condensate would drive the theory into $3d$ topological ordered phase with nontrivial particle-loop(or three-loop) braiding. In particular, such transition could be mapped into the surface state of a $Z_2$ symmetry breaking phase transition in $4d$ where the $Z_2$ domain wall is decorated with $3d$ SPT state. Elucidating this connection provides us a new way to relate the topological phase transition and conventional LGW transition via a holographic view.

Besides exploring the effective theory of the unconventional phase transition, another motivation of this work is to seek exotic phase of matter and their transition in Weyl semimetals. Due to the monopole carried by the Weyl cone in momentum space, a class of boson order parameters who gap out the Weyl semimetal exhibit a WZW term or topological $\Theta$ term. This provides a systematic path for us to explore the novel boson quantum phases via fermion models. We anticipate our approach could shed light on the exploration of exotic phases in Weyl semimetal systems.

\begin{acknowledgements}
We are grateful to S Ryu, Y-Z You, D-T Son, L Zou, X Chen and E Fradkin for insightful comments and discussions. This work was supported in part by the National Science Foundation through grants DMR-1408713 (YY) at the University of Illinois. YY acknowledge support by the 2016 Boulder Summer School for Condensed
Matter and Materials Physics through NSF grant DMR-13001648 when part of this work was done.
\end{acknowledgements}

\providecommand{\noopsort}[1]{}\providecommand{\singleletter}[1]{#1}%


\begin{thebibliography}{54}%
\makeatletter
\providecommand \@ifxundefined [1]{%
 \@ifx{#1\undefined}
}%
\providecommand \@ifnum [1]{%
 \ifnum #1\expandafter \@firstoftwo
 \else \expandafter \@secondoftwo
 \fi
}%
\providecommand \@ifx [1]{%
 \ifx #1\expandafter \@firstoftwo
 \else \expandafter \@secondoftwo
 \fi
}%
\providecommand \natexlab [1]{#1}%
\providecommand \enquote  [1]{``#1''}%
\providecommand \bibnamefont  [1]{#1}%
\providecommand \bibfnamefont [1]{#1}%
\providecommand \citenamefont [1]{#1}%
\providecommand \href@noop [0]{\@secondoftwo}%
\providecommand \href [0]{\begingroup \@sanitize@url \@href}%
\providecommand \@href[1]{\@@startlink{#1}\@@href}%
\providecommand \@@href[1]{\endgroup#1\@@endlink}%
\providecommand \@sanitize@url [0]{\catcode `\\12\catcode `\$12\catcode
  `\&12\catcode `\#12\catcode `\^12\catcode `\_12\catcode `\%12\relax}%
\providecommand \@@startlink[1]{}%
\providecommand \@@endlink[0]{}%
\providecommand \url  [0]{\begingroup\@sanitize@url \@url }%
\providecommand \@url [1]{\endgroup\@href {#1}{\urlprefix }}%
\providecommand \urlprefix  [0]{URL }%
\providecommand \Eprint [0]{\href }%
\providecommand \doibase [0]{http://dx.doi.org/}%
\providecommand \selectlanguage [0]{\@gobble}%
\providecommand \bibinfo  [0]{\@secondoftwo}%
\providecommand \bibfield  [0]{\@secondoftwo}%
\providecommand \translation [1]{[#1]}%
\providecommand \BibitemOpen [0]{}%
\providecommand \bibitemStop [0]{}%
\providecommand \bibitemNoStop [0]{.\EOS\space}%
\providecommand \EOS [0]{\spacefactor3000\relax}%
\providecommand \BibitemShut  [1]{\csname bibitem#1\endcsname}%
\let\auto@bib@innerbib\@empty
\bibitem [{\citenamefont {Pfeuty}(1970)}]{pfeuty1970one}%
  \BibitemOpen
  \bibfield  {author} {\bibinfo {author} {\bibfnamefont {P.}~\bibnamefont
  {Pfeuty}},\ }\href@noop {} {\bibfield  {journal} {\bibinfo  {journal} {ANNALS
  of Physics}\ }\textbf {\bibinfo {volume} {57}},\ \bibinfo {pages} {79}
  (\bibinfo {year} {1970})}\BibitemShut {NoStop}%
\bibitem [{\citenamefont {Fisher}\ \emph {et~al.}(1989)\citenamefont {Fisher},
  \citenamefont {Weichman}, \citenamefont {Grinstein},\ and\ \citenamefont
  {Fisher}}]{fisher1989boson}%
  \BibitemOpen
  \bibfield  {author} {\bibinfo {author} {\bibfnamefont {M.~P.}\ \bibnamefont
  {Fisher}}, \bibinfo {author} {\bibfnamefont {P.~B.}\ \bibnamefont
  {Weichman}}, \bibinfo {author} {\bibfnamefont {G.}~\bibnamefont {Grinstein}},
  \ and\ \bibinfo {author} {\bibfnamefont {D.~S.}\ \bibnamefont {Fisher}},\
  }\href@noop {} {\bibfield  {journal} {\bibinfo  {journal} {Physical Review
  B}\ }\textbf {\bibinfo {volume} {40}},\ \bibinfo {pages} {546} (\bibinfo
  {year} {1989})}\BibitemShut {NoStop}%
\bibitem [{\citenamefont {Lee}\ and\ \citenamefont
  {Kane}(1990)}]{lee1990boson}%
  \BibitemOpen
  \bibfield  {author} {\bibinfo {author} {\bibfnamefont {D.-H.}\ \bibnamefont
  {Lee}}\ and\ \bibinfo {author} {\bibfnamefont {C.~L.}\ \bibnamefont {Kane}},\
  }\href@noop {} {\bibfield  {journal} {\bibinfo  {journal} {Physical review
  letters}\ }\textbf {\bibinfo {volume} {64}},\ \bibinfo {pages} {1313}
  (\bibinfo {year} {1990})}\BibitemShut {NoStop}%
\bibitem [{\citenamefont {Senthil}\ \emph {et~al.}(2005)\citenamefont
  {Senthil}, \citenamefont {Balents}, \citenamefont {Sachdev}, \citenamefont
  {Vishwanath},\ and\ \citenamefont {PA~Fisher}}]{senthil2005deconfined}%
  \BibitemOpen
  \bibfield  {author} {\bibinfo {author} {\bibfnamefont {T.}~\bibnamefont
  {Senthil}}, \bibinfo {author} {\bibfnamefont {L.}~\bibnamefont {Balents}},
  \bibinfo {author} {\bibfnamefont {S.}~\bibnamefont {Sachdev}}, \bibinfo
  {author} {\bibfnamefont {A.}~\bibnamefont {Vishwanath}}, \ and\ \bibinfo
  {author} {\bibfnamefont {M.}~\bibnamefont {PA~Fisher}},\ }\href@noop {}
  {\bibfield  {journal} {\bibinfo  {journal} {Journal of the Physical Society
  of Japan}\ }\textbf {\bibinfo {volume} {74}},\ \bibinfo {pages} {1} (\bibinfo
  {year} {2005})}\BibitemShut {NoStop}%
\bibitem [{\citenamefont {Senthil}\ \emph {et~al.}(2004)\citenamefont
  {Senthil}, \citenamefont {Vishwanath}, \citenamefont {Balents}, \citenamefont
  {Sachdev},\ and\ \citenamefont {Fisher}}]{senthil2004deconfined}%
  \BibitemOpen
  \bibfield  {author} {\bibinfo {author} {\bibfnamefont {T.}~\bibnamefont
  {Senthil}}, \bibinfo {author} {\bibfnamefont {A.}~\bibnamefont {Vishwanath}},
  \bibinfo {author} {\bibfnamefont {L.}~\bibnamefont {Balents}}, \bibinfo
  {author} {\bibfnamefont {S.}~\bibnamefont {Sachdev}}, \ and\ \bibinfo
  {author} {\bibfnamefont {M.~P.}\ \bibnamefont {Fisher}},\ }\href@noop {}
  {\bibfield  {journal} {\bibinfo  {journal} {Science}\ }\textbf {\bibinfo
  {volume} {303}},\ \bibinfo {pages} {1490} (\bibinfo {year}
  {2004})}\BibitemShut {NoStop}%
\bibitem [{\citenamefont {Vishwanath}\ \emph {et~al.}(2004)\citenamefont
  {Vishwanath}, \citenamefont {Balents},\ and\ \citenamefont
  {Senthil}}]{vishwanath2004quantum}%
  \BibitemOpen
  \bibfield  {author} {\bibinfo {author} {\bibfnamefont {A.}~\bibnamefont
  {Vishwanath}}, \bibinfo {author} {\bibfnamefont {L.}~\bibnamefont {Balents}},
  \ and\ \bibinfo {author} {\bibfnamefont {T.}~\bibnamefont {Senthil}},\
  }\href@noop {} {\bibfield  {journal} {\bibinfo  {journal} {Physical Review
  B}\ }\textbf {\bibinfo {volume} {69}},\ \bibinfo {pages} {224416} (\bibinfo
  {year} {2004})}\BibitemShut {NoStop}%
\bibitem [{\citenamefont {Senthil}\ and\ \citenamefont
  {Fisher}(2006)}]{senthil2006competing}%
  \BibitemOpen
  \bibfield  {author} {\bibinfo {author} {\bibfnamefont {T.}~\bibnamefont
  {Senthil}}\ and\ \bibinfo {author} {\bibfnamefont {M.~P.}\ \bibnamefont
  {Fisher}},\ }\href@noop {} {\bibfield  {journal} {\bibinfo  {journal}
  {Physical Review B}\ }\textbf {\bibinfo {volume} {74}},\ \bibinfo {pages}
  {064405} (\bibinfo {year} {2006})}\BibitemShut {NoStop}%
\bibitem [{\citenamefont {Grover}\ and\ \citenamefont
  {Senthil}(2008)}]{grover2008topological}%
  \BibitemOpen
  \bibfield  {author} {\bibinfo {author} {\bibfnamefont {T.}~\bibnamefont
  {Grover}}\ and\ \bibinfo {author} {\bibfnamefont {T.}~\bibnamefont
  {Senthil}},\ }\href@noop {} {\bibfield  {journal} {\bibinfo  {journal}
  {Physical review letters}\ }\textbf {\bibinfo {volume} {100}},\ \bibinfo
  {pages} {156804} (\bibinfo {year} {2008})}\BibitemShut {NoStop}%
\bibitem [{\citenamefont {Moon}(2012)}]{moon2012skyrmions}%
  \BibitemOpen
  \bibfield  {author} {\bibinfo {author} {\bibfnamefont {E.-G.}\ \bibnamefont
  {Moon}},\ }\href@noop {} {\bibfield  {journal} {\bibinfo  {journal} {Physical
  Review B}\ }\textbf {\bibinfo {volume} {85}},\ \bibinfo {pages} {245123}
  (\bibinfo {year} {2012})}\BibitemShut {NoStop}%
\bibitem [{\citenamefont {Tsui}\ \emph
  {et~al.}(2015{\natexlab{a}})\citenamefont {Tsui}, \citenamefont {Jiang},
  \citenamefont {Lu},\ and\ \citenamefont {Lee}}]{tsui2015quantum}%
  \BibitemOpen
  \bibfield  {author} {\bibinfo {author} {\bibfnamefont {L.}~\bibnamefont
  {Tsui}}, \bibinfo {author} {\bibfnamefont {H.-C.}\ \bibnamefont {Jiang}},
  \bibinfo {author} {\bibfnamefont {Y.-M.}\ \bibnamefont {Lu}}, \ and\ \bibinfo
  {author} {\bibfnamefont {D.-H.}\ \bibnamefont {Lee}},\ }\href@noop {}
  {\bibfield  {journal} {\bibinfo  {journal} {Nuclear Physics B}\ }\textbf
  {\bibinfo {volume} {896}},\ \bibinfo {pages} {330} (\bibinfo {year}
  {2015}{\natexlab{a}})}\BibitemShut {NoStop}%
\bibitem [{\citenamefont {Levin}\ and\ \citenamefont
  {Senthil}(2004)}]{levin2004deconfined}%
  \BibitemOpen
  \bibfield  {author} {\bibinfo {author} {\bibfnamefont {M.}~\bibnamefont
  {Levin}}\ and\ \bibinfo {author} {\bibfnamefont {T.}~\bibnamefont
  {Senthil}},\ }\href@noop {} {\bibfield  {journal} {\bibinfo  {journal}
  {Physical Review B}\ }\textbf {\bibinfo {volume} {70}},\ \bibinfo {pages}
  {220403} (\bibinfo {year} {2004})}\BibitemShut {NoStop}%
\bibitem [{\citenamefont {Nahum}\ \emph {et~al.}(2015)\citenamefont {Nahum},
  \citenamefont {Serna}, \citenamefont {Chalker}, \citenamefont {Ortu{\~n}o},\
  and\ \citenamefont {Somoza}}]{nahum2015emergent}%
  \BibitemOpen
  \bibfield  {author} {\bibinfo {author} {\bibfnamefont {A.}~\bibnamefont
  {Nahum}}, \bibinfo {author} {\bibfnamefont {P.}~\bibnamefont {Serna}},
  \bibinfo {author} {\bibfnamefont {J.}~\bibnamefont {Chalker}}, \bibinfo
  {author} {\bibfnamefont {M.}~\bibnamefont {Ortu{\~n}o}}, \ and\ \bibinfo
  {author} {\bibfnamefont {A.}~\bibnamefont {Somoza}},\ }\href@noop {}
  {\bibfield  {journal} {\bibinfo  {journal} {Physical review letters}\
  }\textbf {\bibinfo {volume} {115}},\ \bibinfo {pages} {267203} (\bibinfo
  {year} {2015})}\BibitemShut {NoStop}%
\bibitem [{\citenamefont {Wen}(1990)}]{wen1990topological}%
  \BibitemOpen
  \bibfield  {author} {\bibinfo {author} {\bibfnamefont {X.-G.}\ \bibnamefont
  {Wen}},\ }\href@noop {} {\bibfield  {journal} {\bibinfo  {journal}
  {International Journal of Modern Physics B}\ }\textbf {\bibinfo {volume}
  {4}},\ \bibinfo {pages} {239} (\bibinfo {year} {1990})}\BibitemShut {NoStop}%
\bibitem [{\citenamefont {Laughlin}(1983)}]{laughlin1983anomalous}%
  \BibitemOpen
  \bibfield  {author} {\bibinfo {author} {\bibfnamefont {R.~B.}\ \bibnamefont
  {Laughlin}},\ }\href@noop {} {\bibfield  {journal} {\bibinfo  {journal}
  {Physical Review Letters}\ }\textbf {\bibinfo {volume} {50}},\ \bibinfo
  {pages} {1395} (\bibinfo {year} {1983})}\BibitemShut {NoStop}%
\bibitem [{\citenamefont {Burnell}\ \emph {et~al.}(2012)\citenamefont
  {Burnell}, \citenamefont {Simon},\ and\ \citenamefont
  {Slingerland}}]{burnell2012phase}%
  \BibitemOpen
  \bibfield  {author} {\bibinfo {author} {\bibfnamefont {F.}~\bibnamefont
  {Burnell}}, \bibinfo {author} {\bibfnamefont {S.~H.}\ \bibnamefont {Simon}},
  \ and\ \bibinfo {author} {\bibfnamefont {J.}~\bibnamefont {Slingerland}},\
  }\href@noop {} {\bibfield  {journal} {\bibinfo  {journal} {New Journal of
  Physics}\ }\textbf {\bibinfo {volume} {14}},\ \bibinfo {pages} {015004}
  (\bibinfo {year} {2012})}\BibitemShut {NoStop}%
\bibitem [{\citenamefont {Fisher}(1990)}]{fisher1990quantum}%
  \BibitemOpen
  \bibfield  {author} {\bibinfo {author} {\bibfnamefont {M.~P.}\ \bibnamefont
  {Fisher}},\ }\href@noop {} {\bibfield  {journal} {\bibinfo  {journal}
  {Physical Review Letters}\ }\textbf {\bibinfo {volume} {65}},\ \bibinfo
  {pages} {923} (\bibinfo {year} {1990})}\BibitemShut {NoStop}%
\bibitem [{\citenamefont {Xu}(2012)}]{xu2012unconventional}%
  \BibitemOpen
  \bibfield  {author} {\bibinfo {author} {\bibfnamefont {C.}~\bibnamefont
  {Xu}},\ }\href@noop {} {\bibfield  {journal} {\bibinfo  {journal}
  {International Journal of Modern Physics B}\ }\textbf {\bibinfo {volume}
  {26}},\ \bibinfo {pages} {1230007} (\bibinfo {year} {2012})}\BibitemShut
  {NoStop}%
\bibitem [{\citenamefont {Lee}\ and\ \citenamefont
  {FISHER}(1991)}]{lee1991anyon}%
  \BibitemOpen
  \bibfield  {author} {\bibinfo {author} {\bibfnamefont {D.-H.}\ \bibnamefont
  {Lee}}\ and\ \bibinfo {author} {\bibfnamefont {M.~P.}\ \bibnamefont
  {FISHER}},\ }\href@noop {} {\bibfield  {journal} {\bibinfo  {journal}
  {International Journal of Modern Physics B}\ }\textbf {\bibinfo {volume}
  {5}},\ \bibinfo {pages} {2675} (\bibinfo {year} {1991})}\BibitemShut
  {NoStop}%
\bibitem [{\citenamefont {Senthil}(2014)}]{senthil2014symmetry}%
  \BibitemOpen
  \bibfield  {author} {\bibinfo {author} {\bibfnamefont {T.}~\bibnamefont
  {Senthil}},\ }\href@noop {} {\bibfield  {journal} {\bibinfo  {journal} {arXiv
  preprint arXiv:1405.4015}\ } (\bibinfo {year} {2014})}\BibitemShut {NoStop}%
\bibitem [{\citenamefont {Chen}\ \emph {et~al.}(2014)\citenamefont {Chen},
  \citenamefont {Lu},\ and\ \citenamefont {Vishwanath}}]{chen2014symmetry}%
  \BibitemOpen
  \bibfield  {author} {\bibinfo {author} {\bibfnamefont {X.}~\bibnamefont
  {Chen}}, \bibinfo {author} {\bibfnamefont {Y.-M.}\ \bibnamefont {Lu}}, \ and\
  \bibinfo {author} {\bibfnamefont {A.}~\bibnamefont {Vishwanath}},\
  }\href@noop {} {\bibfield  {journal} {\bibinfo  {journal} {Nature
  communications}\ }\textbf {\bibinfo {volume} {5}} (\bibinfo {year}
  {2014})}\BibitemShut {NoStop}%
\bibitem [{\citenamefont {You}\ and\ \citenamefont
  {Xu}(2014)}]{you2014symmetry}%
  \BibitemOpen
  \bibfield  {author} {\bibinfo {author} {\bibfnamefont {Y.-Z.}\ \bibnamefont
  {You}}\ and\ \bibinfo {author} {\bibfnamefont {C.}~\bibnamefont {Xu}},\
  }\href@noop {} {\bibfield  {journal} {\bibinfo  {journal} {Physical Review
  B}\ }\textbf {\bibinfo {volume} {90}},\ \bibinfo {pages} {245120} (\bibinfo
  {year} {2014})}\BibitemShut {NoStop}%
\bibitem [{\citenamefont {Fidkowski}\ and\ \citenamefont
  {Kitaev}(2011)}]{fidkowski2011topological}%
  \BibitemOpen
  \bibfield  {author} {\bibinfo {author} {\bibfnamefont {L.}~\bibnamefont
  {Fidkowski}}\ and\ \bibinfo {author} {\bibfnamefont {A.}~\bibnamefont
  {Kitaev}},\ }\href@noop {} {\bibfield  {journal} {\bibinfo  {journal}
  {Physical Review B}\ }\textbf {\bibinfo {volume} {83}},\ \bibinfo {pages}
  {075103} (\bibinfo {year} {2011})}\BibitemShut {NoStop}%
\bibitem [{\citenamefont {Moon}\ and\ \citenamefont
  {Xu}(2012)}]{moon2012exotic}%
  \BibitemOpen
  \bibfield  {author} {\bibinfo {author} {\bibfnamefont {E.-G.}\ \bibnamefont
  {Moon}}\ and\ \bibinfo {author} {\bibfnamefont {C.}~\bibnamefont {Xu}},\
  }\href@noop {} {\bibfield  {journal} {\bibinfo  {journal} {Physical Review
  B}\ }\textbf {\bibinfo {volume} {86}},\ \bibinfo {pages} {214414} (\bibinfo
  {year} {2012})}\BibitemShut {NoStop}%
\bibitem [{\citenamefont {Kong}(2014)}]{kong2014anyon}%
  \BibitemOpen
  \bibfield  {author} {\bibinfo {author} {\bibfnamefont {L.}~\bibnamefont
  {Kong}},\ }\href@noop {} {\bibfield  {journal} {\bibinfo  {journal} {Nuclear
  Physics B}\ }\textbf {\bibinfo {volume} {886}},\ \bibinfo {pages} {436}
  (\bibinfo {year} {2014})}\BibitemShut {NoStop}%
\bibitem [{\citenamefont {Barkeshli}\ and\ \citenamefont
  {Wen}(2010)}]{barkeshli2010anyon}%
  \BibitemOpen
  \bibfield  {author} {\bibinfo {author} {\bibfnamefont {M.}~\bibnamefont
  {Barkeshli}}\ and\ \bibinfo {author} {\bibfnamefont {X.-G.}\ \bibnamefont
  {Wen}},\ }\href@noop {} {\bibfield  {journal} {\bibinfo  {journal} {Physical
  review letters}\ }\textbf {\bibinfo {volume} {105}},\ \bibinfo {pages}
  {216804} (\bibinfo {year} {2010})}\BibitemShut {NoStop}%
\bibitem [{\citenamefont {Metlitski}\ \emph {et~al.}(2014)\citenamefont
  {Metlitski}, \citenamefont {Fidkowski}, \citenamefont {Chen},\ and\
  \citenamefont {Vishwanath}}]{metlitski2014interaction}%
  \BibitemOpen
  \bibfield  {author} {\bibinfo {author} {\bibfnamefont {M.~A.}\ \bibnamefont
  {Metlitski}}, \bibinfo {author} {\bibfnamefont {L.}~\bibnamefont
  {Fidkowski}}, \bibinfo {author} {\bibfnamefont {X.}~\bibnamefont {Chen}}, \
  and\ \bibinfo {author} {\bibfnamefont {A.}~\bibnamefont {Vishwanath}},\
  }\href@noop {} {\bibfield  {journal} {\bibinfo  {journal} {arXiv preprint
  arXiv:1406.3032}\ } (\bibinfo {year} {2014})}\BibitemShut {NoStop}%
\bibitem [{\citenamefont {You}\ and\ \citenamefont
  {You}(2016{\natexlab{a}})}]{you2016stripe}%
  \BibitemOpen
  \bibfield  {author} {\bibinfo {author} {\bibfnamefont {Y.}~\bibnamefont
  {You}}\ and\ \bibinfo {author} {\bibfnamefont {Y.-Z.}\ \bibnamefont {You}},\
  }\href@noop {} {\bibfield  {journal} {\bibinfo  {journal} {Physical Review
  B}\ }\textbf {\bibinfo {volume} {93}},\ \bibinfo {pages} {195141} (\bibinfo
  {year} {2016}{\natexlab{a}})}\BibitemShut {NoStop}%
\bibitem [{\citenamefont {You}\ and\ \citenamefont
  {You}(2016{\natexlab{b}})}]{you2016geometry}%
  \BibitemOpen
  \bibfield  {author} {\bibinfo {author} {\bibfnamefont {Y.}~\bibnamefont
  {You}}\ and\ \bibinfo {author} {\bibfnamefont {Y.-Z.}\ \bibnamefont {You}},\
  }\href@noop {} {\bibfield  {journal} {\bibinfo  {journal} {arXiv preprint
  arXiv:1603.02694}\ } (\bibinfo {year} {2016}{\natexlab{b}})}\BibitemShut
  {NoStop}%
\bibitem [{\citenamefont {Cho}\ \emph {et~al.}(2015)\citenamefont {Cho},
  \citenamefont {Parrikar}, \citenamefont {You}, \citenamefont {Leigh},\ and\
  \citenamefont {Hughes}}]{cho2015condensation}%
  \BibitemOpen
  \bibfield  {author} {\bibinfo {author} {\bibfnamefont {G.~Y.}\ \bibnamefont
  {Cho}}, \bibinfo {author} {\bibfnamefont {O.}~\bibnamefont {Parrikar}},
  \bibinfo {author} {\bibfnamefont {Y.}~\bibnamefont {You}}, \bibinfo {author}
  {\bibfnamefont {R.~G.}\ \bibnamefont {Leigh}}, \ and\ \bibinfo {author}
  {\bibfnamefont {T.~L.}\ \bibnamefont {Hughes}},\ }\href@noop {} {\bibfield
  {journal} {\bibinfo  {journal} {Physical Review B}\ }\textbf {\bibinfo
  {volume} {91}},\ \bibinfo {pages} {035122} (\bibinfo {year}
  {2015})}\BibitemShut {NoStop}%
\bibitem [{\citenamefont {Chen}\ \emph {et~al.}(2013)\citenamefont {Chen},
  \citenamefont {Wang}, \citenamefont {Lu},\ and\ \citenamefont
  {Lee}}]{chen2013critical}%
  \BibitemOpen
  \bibfield  {author} {\bibinfo {author} {\bibfnamefont {X.}~\bibnamefont
  {Chen}}, \bibinfo {author} {\bibfnamefont {F.}~\bibnamefont {Wang}}, \bibinfo
  {author} {\bibfnamefont {Y.-M.}\ \bibnamefont {Lu}}, \ and\ \bibinfo {author}
  {\bibfnamefont {D.-H.}\ \bibnamefont {Lee}},\ }\href@noop {} {\bibfield
  {journal} {\bibinfo  {journal} {Nuclear Physics B}\ }\textbf {\bibinfo
  {volume} {873}},\ \bibinfo {pages} {248} (\bibinfo {year}
  {2013})}\BibitemShut {NoStop}%
\bibitem [{\citenamefont {Tsui}\ \emph
  {et~al.}(2015{\natexlab{b}})\citenamefont {Tsui}, \citenamefont {Wang},\ and\
  \citenamefont {Lee}}]{tsui2015topological}%
  \BibitemOpen
  \bibfield  {author} {\bibinfo {author} {\bibfnamefont {L.}~\bibnamefont
  {Tsui}}, \bibinfo {author} {\bibfnamefont {F.}~\bibnamefont {Wang}}, \ and\
  \bibinfo {author} {\bibfnamefont {D.-H.}\ \bibnamefont {Lee}},\ }\href@noop
  {} {\bibfield  {journal} {\bibinfo  {journal} {arXiv preprint
  arXiv:1511.07460}\ } (\bibinfo {year} {2015}{\natexlab{b}})}\BibitemShut
  {NoStop}%
\bibitem [{\citenamefont {Wan}\ \emph {et~al.}(2011)\citenamefont {Wan},
  \citenamefont {Turner}, \citenamefont {Vishwanath},\ and\ \citenamefont
  {Savrasov}}]{wan2011topological}%
  \BibitemOpen
  \bibfield  {author} {\bibinfo {author} {\bibfnamefont {X.}~\bibnamefont
  {Wan}}, \bibinfo {author} {\bibfnamefont {A.~M.}\ \bibnamefont {Turner}},
  \bibinfo {author} {\bibfnamefont {A.}~\bibnamefont {Vishwanath}}, \ and\
  \bibinfo {author} {\bibfnamefont {S.~Y.}\ \bibnamefont {Savrasov}},\
  }\href@noop {} {\bibfield  {journal} {\bibinfo  {journal} {Physical Review
  B}\ }\textbf {\bibinfo {volume} {83}},\ \bibinfo {pages} {205101} (\bibinfo
  {year} {2011})}\BibitemShut {NoStop}%
\bibitem [{\citenamefont {Zyuzin}\ and\ \citenamefont
  {Burkov}(2012)}]{zyuzin2012topological}%
  \BibitemOpen
  \bibfield  {author} {\bibinfo {author} {\bibfnamefont {A.}~\bibnamefont
  {Zyuzin}}\ and\ \bibinfo {author} {\bibfnamefont {A.}~\bibnamefont
  {Burkov}},\ }\href@noop {} {\bibfield  {journal} {\bibinfo  {journal}
  {Physical Review B}\ }\textbf {\bibinfo {volume} {86}},\ \bibinfo {pages}
  {115133} (\bibinfo {year} {2012})}\BibitemShut {NoStop}%
\bibitem [{\citenamefont {Yu}\ \emph {et~al.}(2008)\citenamefont {Yu},
  \citenamefont {Kou},\ and\ \citenamefont {Wen}}]{yu2008topological}%
  \BibitemOpen
  \bibfield  {author} {\bibinfo {author} {\bibfnamefont {J.}~\bibnamefont
  {Yu}}, \bibinfo {author} {\bibfnamefont {S.-P.}\ \bibnamefont {Kou}}, \ and\
  \bibinfo {author} {\bibfnamefont {X.-G.}\ \bibnamefont {Wen}},\ }\href@noop
  {} {\bibfield  {journal} {\bibinfo  {journal} {EPL (Europhysics Letters)}\
  }\textbf {\bibinfo {volume} {84}},\ \bibinfo {pages} {17004} (\bibinfo {year}
  {2008})}\BibitemShut {NoStop}%
\bibitem [{\citenamefont {Moradi}\ and\ \citenamefont
  {Wen}(2015)}]{moradi2015universal}%
  \BibitemOpen
  \bibfield  {author} {\bibinfo {author} {\bibfnamefont {H.}~\bibnamefont
  {Moradi}}\ and\ \bibinfo {author} {\bibfnamefont {X.-G.}\ \bibnamefont
  {Wen}},\ }\href@noop {} {\bibfield  {journal} {\bibinfo  {journal} {Physical
  Review B}\ }\textbf {\bibinfo {volume} {91}},\ \bibinfo {pages} {075114}
  (\bibinfo {year} {2015})}\BibitemShut {NoStop}%
\bibitem [{\citenamefont {Haldane}(1983)}]{haldane1983continuum}%
  \BibitemOpen
  \bibfield  {author} {\bibinfo {author} {\bibfnamefont {F.~D.~M.}\
  \bibnamefont {Haldane}},\ }\href@noop {} {\bibfield  {journal} {\bibinfo
  {journal} {Physics Letters A}\ }\textbf {\bibinfo {volume} {93}},\ \bibinfo
  {pages} {464} (\bibinfo {year} {1983})}\BibitemShut {NoStop}%
\bibitem [{\citenamefont {Moon}(2015)}]{moon2015competing}%
  \BibitemOpen
  \bibfield  {author} {\bibinfo {author} {\bibfnamefont {E.-G.}\ \bibnamefont
  {Moon}},\ }\href@noop {} {\bibfield  {journal} {\bibinfo  {journal} {arXiv
  preprint arXiv:1503.05199}\ } (\bibinfo {year} {2015})}\BibitemShut {NoStop}%
\bibitem [{\citenamefont {Xu}\ and\ \citenamefont
  {Ludwig}(2013)}]{xu2013nonperturbative}%
  \BibitemOpen
  \bibfield  {author} {\bibinfo {author} {\bibfnamefont {C.}~\bibnamefont
  {Xu}}\ and\ \bibinfo {author} {\bibfnamefont {A.~W.}\ \bibnamefont
  {Ludwig}},\ }\href@noop {} {\bibfield  {journal} {\bibinfo  {journal}
  {Physical review letters}\ }\textbf {\bibinfo {volume} {110}},\ \bibinfo
  {pages} {200405} (\bibinfo {year} {2013})}\BibitemShut {NoStop}%
\bibitem [{\citenamefont {Chan}\ \emph {et~al.}(2016)\citenamefont {Chan},
  \citenamefont {Kvorning}, \citenamefont {Ryu},\ and\ \citenamefont
  {Fradkin}}]{chan2016effective}%
  \BibitemOpen
  \bibfield  {author} {\bibinfo {author} {\bibfnamefont {A.~P.}\ \bibnamefont
  {Chan}}, \bibinfo {author} {\bibfnamefont {T.}~\bibnamefont {Kvorning}},
  \bibinfo {author} {\bibfnamefont {S.}~\bibnamefont {Ryu}}, \ and\ \bibinfo
  {author} {\bibfnamefont {E.}~\bibnamefont {Fradkin}},\ }\href@noop {}
  {\bibfield  {journal} {\bibinfo  {journal} {Physical Review B}\ }\textbf
  {\bibinfo {volume} {93}},\ \bibinfo {pages} {155122} (\bibinfo {year}
  {2016})}\BibitemShut {NoStop}%
\bibitem [{\citenamefont {Xu}\ and\ \citenamefont
  {Senthil}(2013)}]{xu2013wave}%
  \BibitemOpen
  \bibfield  {author} {\bibinfo {author} {\bibfnamefont {C.}~\bibnamefont
  {Xu}}\ and\ \bibinfo {author} {\bibfnamefont {T.}~\bibnamefont {Senthil}},\
  }\href@noop {} {\bibfield  {journal} {\bibinfo  {journal} {Physical Review
  B}\ }\textbf {\bibinfo {volume} {87}},\ \bibinfo {pages} {174412} (\bibinfo
  {year} {2013})}\BibitemShut {NoStop}%
\bibitem [{\citenamefont {Bi}\ \emph {et~al.}(2014{\natexlab{a}})\citenamefont
  {Bi}, \citenamefont {Rasmussen}, \citenamefont {You}, \citenamefont {Cheng},\
  and\ \citenamefont {Xu}}]{bi2014bridging}%
  \BibitemOpen
  \bibfield  {author} {\bibinfo {author} {\bibfnamefont {Z.}~\bibnamefont
  {Bi}}, \bibinfo {author} {\bibfnamefont {A.}~\bibnamefont {Rasmussen}},
  \bibinfo {author} {\bibfnamefont {Y.-Z.}\ \bibnamefont {You}}, \bibinfo
  {author} {\bibfnamefont {M.}~\bibnamefont {Cheng}}, \ and\ \bibinfo {author}
  {\bibfnamefont {C.}~\bibnamefont {Xu}},\ }\href@noop {} {\bibfield  {journal}
  {\bibinfo  {journal} {arXiv preprint arXiv:1404.6256}\ } (\bibinfo {year}
  {2014}{\natexlab{a}})}\BibitemShut {NoStop}%
\bibitem [{\citenamefont {Abanov}\ and\ \citenamefont
  {Wiegmann}(2000)}]{abanov2000theta}%
  \BibitemOpen
  \bibfield  {author} {\bibinfo {author} {\bibfnamefont {A.}~\bibnamefont
  {Abanov}}\ and\ \bibinfo {author} {\bibfnamefont {P.~B.}\ \bibnamefont
  {Wiegmann}},\ }\href@noop {} {\bibfield  {journal} {\bibinfo  {journal}
  {Nuclear Physics B}\ }\textbf {\bibinfo {volume} {570}},\ \bibinfo {pages}
  {685} (\bibinfo {year} {2000})}\BibitemShut {NoStop}%
\bibitem [{\citenamefont {You}\ and\ \citenamefont
  {Xu}(2015)}]{you2015topological}%
  \BibitemOpen
  \bibfield  {author} {\bibinfo {author} {\bibfnamefont {Y.-Z.}\ \bibnamefont
  {You}}\ and\ \bibinfo {author} {\bibfnamefont {C.}~\bibnamefont {Xu}},\
  }\href@noop {} {\bibfield  {journal} {\bibinfo  {journal} {Physical Review
  B}\ }\textbf {\bibinfo {volume} {92}},\ \bibinfo {pages} {054410} (\bibinfo
  {year} {2015})}\BibitemShut {NoStop}%
\bibitem [{\citenamefont {Ran}\ \emph {et~al.}(2011)\citenamefont {Ran},
  \citenamefont {Hosur},\ and\ \citenamefont {Vishwanath}}]{ran2011fermionic}%
  \BibitemOpen
  \bibfield  {author} {\bibinfo {author} {\bibfnamefont {Y.}~\bibnamefont
  {Ran}}, \bibinfo {author} {\bibfnamefont {P.}~\bibnamefont {Hosur}}, \ and\
  \bibinfo {author} {\bibfnamefont {A.}~\bibnamefont {Vishwanath}},\
  }\href@noop {} {\bibfield  {journal} {\bibinfo  {journal} {Physical Review
  B}\ }\textbf {\bibinfo {volume} {84}},\ \bibinfo {pages} {184501} (\bibinfo
  {year} {2011})}\BibitemShut {NoStop}%
\bibitem [{\citenamefont {Bi}\ \emph {et~al.}(2015{\natexlab{a}})\citenamefont
  {Bi}, \citenamefont {Slagle},\ and\ \citenamefont {Xu}}]{bi2015self}%
  \BibitemOpen
  \bibfield  {author} {\bibinfo {author} {\bibfnamefont {Z.}~\bibnamefont
  {Bi}}, \bibinfo {author} {\bibfnamefont {K.}~\bibnamefont {Slagle}}, \ and\
  \bibinfo {author} {\bibfnamefont {C.}~\bibnamefont {Xu}},\ }\href@noop {}
  {\bibfield  {journal} {\bibinfo  {journal} {arXiv preprint arXiv:1504.04373}\
  } (\bibinfo {year} {2015}{\natexlab{a}})}\BibitemShut {NoStop}%
\bibitem [{\citenamefont {Balents}\ \emph {et~al.}(1999)\citenamefont
  {Balents}, \citenamefont {Fisher},\ and\ \citenamefont
  {Nayak}}]{balents1999dual}%
  \BibitemOpen
  \bibfield  {author} {\bibinfo {author} {\bibfnamefont {L.}~\bibnamefont
  {Balents}}, \bibinfo {author} {\bibfnamefont {M.~P.}\ \bibnamefont {Fisher}},
  \ and\ \bibinfo {author} {\bibfnamefont {C.}~\bibnamefont {Nayak}},\
  }\href@noop {} {\bibfield  {journal} {\bibinfo  {journal} {Physical Review
  B}\ }\textbf {\bibinfo {volume} {60}},\ \bibinfo {pages} {1654} (\bibinfo
  {year} {1999})}\BibitemShut {NoStop}%
\bibitem [{\citenamefont {Berg}\ \emph {et~al.}(2009)\citenamefont {Berg},
  \citenamefont {Fradkin},\ and\ \citenamefont {Kivelson}}]{berg2009charge}%
  \BibitemOpen
  \bibfield  {author} {\bibinfo {author} {\bibfnamefont {E.}~\bibnamefont
  {Berg}}, \bibinfo {author} {\bibfnamefont {E.}~\bibnamefont {Fradkin}}, \
  and\ \bibinfo {author} {\bibfnamefont {S.~A.}\ \bibnamefont {Kivelson}},\
  }\href@noop {} {\bibfield  {journal} {\bibinfo  {journal} {Nature Physics}\
  }\textbf {\bibinfo {volume} {5}},\ \bibinfo {pages} {830} (\bibinfo {year}
  {2009})}\BibitemShut {NoStop}%
\bibitem [{\citenamefont {Levin}\ and\ \citenamefont
  {Gu}(2012)}]{levin2012braiding}%
  \BibitemOpen
  \bibfield  {author} {\bibinfo {author} {\bibfnamefont {M.}~\bibnamefont
  {Levin}}\ and\ \bibinfo {author} {\bibfnamefont {Z.-C.}\ \bibnamefont {Gu}},\
  }\href@noop {} {\bibfield  {journal} {\bibinfo  {journal} {Physical Review
  B}\ }\textbf {\bibinfo {volume} {86}},\ \bibinfo {pages} {115109} (\bibinfo
  {year} {2012})}\BibitemShut {NoStop}%
\bibitem [{\citenamefont {Qi}\ \emph {et~al.}(2008)\citenamefont {Qi},
  \citenamefont {Hughes},\ and\ \citenamefont {Zhang}}]{qi2008topological}%
  \BibitemOpen
  \bibfield  {author} {\bibinfo {author} {\bibfnamefont {X.-L.}\ \bibnamefont
  {Qi}}, \bibinfo {author} {\bibfnamefont {T.~L.}\ \bibnamefont {Hughes}}, \
  and\ \bibinfo {author} {\bibfnamefont {S.-C.}\ \bibnamefont {Zhang}},\
  }\href@noop {} {\bibfield  {journal} {\bibinfo  {journal} {Physical Review
  B}\ }\textbf {\bibinfo {volume} {78}},\ \bibinfo {pages} {195424} (\bibinfo
  {year} {2008})}\BibitemShut {NoStop}%
\bibitem [{\citenamefont {Bi}\ \emph {et~al.}(2015{\natexlab{b}})\citenamefont
  {Bi}, \citenamefont {Rasmussen}, \citenamefont {Slagle},\ and\ \citenamefont
  {Xu}}]{bi2015classification}%
  \BibitemOpen
  \bibfield  {author} {\bibinfo {author} {\bibfnamefont {Z.}~\bibnamefont
  {Bi}}, \bibinfo {author} {\bibfnamefont {A.}~\bibnamefont {Rasmussen}},
  \bibinfo {author} {\bibfnamefont {K.}~\bibnamefont {Slagle}}, \ and\ \bibinfo
  {author} {\bibfnamefont {C.}~\bibnamefont {Xu}},\ }\href@noop {} {\bibfield
  {journal} {\bibinfo  {journal} {Physical Review B}\ }\textbf {\bibinfo
  {volume} {91}},\ \bibinfo {pages} {134404} (\bibinfo {year}
  {2015}{\natexlab{b}})}\BibitemShut {NoStop}%
\bibitem [{\citenamefont {Wang}\ and\ \citenamefont
  {Levin}(2014)}]{wang2014braiding}%
  \BibitemOpen
  \bibfield  {author} {\bibinfo {author} {\bibfnamefont {C.}~\bibnamefont
  {Wang}}\ and\ \bibinfo {author} {\bibfnamefont {M.}~\bibnamefont {Levin}},\
  }\href@noop {} {\bibfield  {journal} {\bibinfo  {journal} {Physical review
  letters}\ }\textbf {\bibinfo {volume} {113}},\ \bibinfo {pages} {080403}
  (\bibinfo {year} {2014})}\BibitemShut {NoStop}%
\bibitem [{\citenamefont {Bi}\ \emph {et~al.}(2014{\natexlab{b}})\citenamefont
  {Bi}, \citenamefont {You},\ and\ \citenamefont {Xu}}]{bi2014anyon}%
  \BibitemOpen
  \bibfield  {author} {\bibinfo {author} {\bibfnamefont {Z.}~\bibnamefont
  {Bi}}, \bibinfo {author} {\bibfnamefont {Y.-Z.}\ \bibnamefont {You}}, \ and\
  \bibinfo {author} {\bibfnamefont {C.}~\bibnamefont {Xu}},\ }\href@noop {}
  {\bibfield  {journal} {\bibinfo  {journal} {Physical Review B}\ }\textbf
  {\bibinfo {volume} {90}},\ \bibinfo {pages} {081110} (\bibinfo {year}
  {2014}{\natexlab{b}})}\BibitemShut {NoStop}%
\bibitem [{\citenamefont {Wang}\ and\ \citenamefont
  {Senthil}(2014)}]{wang2014interacting}%
  \BibitemOpen
  \bibfield  {author} {\bibinfo {author} {\bibfnamefont {C.}~\bibnamefont
  {Wang}}\ and\ \bibinfo {author} {\bibfnamefont {T.}~\bibnamefont {Senthil}},\
  }\href@noop {} {\bibfield  {journal} {\bibinfo  {journal} {Physical Review
  B}\ }\textbf {\bibinfo {volume} {89}},\ \bibinfo {pages} {195124} (\bibinfo
  {year} {2014})}\BibitemShut {NoStop}%
\bibitem [{\citenamefont {Jian}\ and\ \citenamefont
  {Qi}(2014)}]{jian2014layer}%
  \BibitemOpen
  \bibfield  {author} {\bibinfo {author} {\bibfnamefont {C.-M.}\ \bibnamefont
  {Jian}}\ and\ \bibinfo {author} {\bibfnamefont {X.-L.}\ \bibnamefont {Qi}},\
  }\href@noop {} {\bibfield  {journal} {\bibinfo  {journal} {Physical Review
  X}\ }\textbf {\bibinfo {volume} {4}},\ \bibinfo {pages} {041043} (\bibinfo
  {year} {2014})}\BibitemShut {NoStop}%
\end{thebibliography}
\end{document}